\documentclass[a4paper,11pt]{article}

\pdfoutput=1
\usepackage{subfigure}
\usepackage{booktabs}

\usepackage{lineno}
\usepackage[colorlinks=false]{hyperref}
\usepackage{bm}
\usepackage{graphicx}
\usepackage{amssymb}
\usepackage{amsmath}
\usepackage{tabularx}
\usepackage{floatrow}
\usepackage{multirow}
\usepackage{caption}
\usepackage{amsmath}
\usepackage{amsfonts}
\usepackage{amssymb}

\newcommand*{\abs}[1]{\left|{#1}\right|}
\newcommand*{\bracket}[1]{\left({#1}\right) } 
\newcommand*{\sbracket}[1]{\left[ {#1}\right] } 

\modulolinenumbers[5]

\usepackage{authblk}

\usepackage[left=25mm,right=25mm,top=1.5cm,bottom=1.5cm,includeheadfoot]{geometry}
\begin{document}

\title{Surrogate modeling based on resampled polynomial chaos expansions}


\author[1]{Zicheng Liu}
\author[2]{Dominique Lesselier}
\author[3]{Bruno Sudret}
\author[1]{Joe Wiart}

\affil[1]{\scriptsize Chaire C2M, LTCI, T\'el\'ecom Paris, Palaiseau 91120, France.}
\affil[2]{\scriptsize Universit\'e Paris-Saclay, CNRS, CentraleSup\'elec,  Laboratoire des signaux et syst\`emes, Gif-sur-Yvette 91190, France}
\affil[3]{\scriptsize ETH Z\"urich, Chair of Risk, Safety and Uncertainty Quantification, Stefano-Franscini-Platz 5, Z\"urich 8093, Switzerland}

\maketitle

\abstract{
In surrogate modeling, polynomial chaos expansion (PCE) is popularly utilized to represent the random model responses, which are computationally expensive and usually obtained by deterministic numerical modeling approaches including finite-element and finite-difference time-domain methods. Recently, efforts have been made on improving the prediction performance of the PCE-based model and building efficiency by only selecting the influential basis polynomials (e.g., via the approach of least angle regression). This paper proposes an approach, named as resampled PCE (rPCE), to further optimize the selection by making use of the knowledge that the true model is fixed despite the statistical uncertainty inherent to sampling in the training. By simulating data variation via resampling ($k$-fold division utilized here) and collecting the selected polynomials with respect to all resamples, polynomials are ranked mainly according to the selection frequency. The resampling scheme (the value of $k$ here) matters much and various configurations are considered and compared. The proposed resampled PCE is implemented with two popular selection techniques, namely least angle regression and orthogonal matching pursuit, and a combination thereof. The performance of the proposed algorithm is demonstrated on one analytical examples, a benchmark problem in structural mechanics, as well as a realistic case study in computational dosimetry.
\\

\noindent
\begin{tabular}{ll}
\textbf{Keywords:} &
Surrogate Modeling, Sparse Polynomial Chaos Expansion, Resampled\\& Polynomial Chaos Expansion, Data Resampling, Sensitivity Analysis,\\& Double Cross Validation
\end{tabular}
}

\section{Introduction}
\label{sec:Introduction}
Mathematical modeling is common practice nowadays for better understanding real-world phenomena. However, a closed-form solution of the governing equations is unavailable in general and numerical modeling schemes, such as finite-difference time-domain (FDTD) \cite{taflove2005computational} and finite element method (FEM) \cite{bathe1976numerical}, are commonly employed. The computational method can be considered as a black-box code that takes a vector of parameters as input and yields a vector of quantities of interest that can be further used to assess the system under consideration. However, the real-world system may not be accurately modeled, one critical factor being the uncertainty of input parameters \cite{laroquetutorial}, which can be taken into account by setting a probabilistic model of these parameters.  

Describing inputs by random variables which follow specific probabilistic density functions (PDFs) \cite{kolmogorov2018foundations}, the propagation of such random inputs through the system yields random outputs and the investigation of such uncertainty propagation is {one of the major problems in} uncertainty quantification (UQ) \cite{iman1988investigation}. Monte Carlo simulations (MCS) can be applied/used to run the UQ analysis, however, it becomes intractable when the computational cost of a single simulation is high (which corresponds with the cases {described} here). Surrogate model (a.k.a. metamodel) is popularly utilized as a remedy to emulate the system response. Among various approaches, such as Gaussian process (Kriging method) \cite{kleijnen2009kriging}, neural networks \cite{mackay1992bayesian}, etc., surrogate modeling based on polynomial chaos expansion (PCE) {\cite{sudret2007uncertainty,sepahvand2010uncertainty,kersaudy2014stochastic,soize2004physical,xiu2002wiener}} is of interest here due to its advantages in both interpretation and versatility.

Representing the finite-variance random output on a Hilbert space spanned by multivariate basis polynomials orthogonal to the joint PDF of input variables, the numerical modeling of the system response is replaced by the computation of a PCE, while the expansion coefficients can be obtained by two different methodologies. For the so-called intrusive methods, taking the spectral finite element method \cite{roger2003stochastic} as an example, the classical FEM is combined with the Karhunen-Lo\`eve expansion of input random fields and the coefficients are obtained by a Galerkin scheme which results in a system of deterministic equations \cite{sudret2004stochastic}. In contrast, without modifying the underlying code, hence as non-intrusive methods, coefficients can be obtained based on an experimental design (ED) by two popularly utilized approaches. While minimizing the mean square error of data discrepancy leads to the solution of regression method \cite{berveiller2006stochastic}, projection method \cite{le2002stochastic,gilli2013uncertainty} exploits the orthogonality of basis functions, the expansion coefficient being the solution of multidimensional integrations which can computed by quadrature methods. 

A PCE, as an infinite series, should be truncated for computational purpose. How to perform this truncation optimally is the major issue, which is addressed in this paper. In the literature, a maximum value is commonly set to the total degree of multivariate polynomials \cite{blatman2009adaptive}. However, the number of basis polynomials, as well as the required ED size, dramatically increases with the number of input variables, which is known as the curse-of-dimensionality. Thus, the so-called sparse PCE \cite{blatman2010adaptive,blatman2011adaptive,doostan2011non,jakeman2015enhancing} has been developed by only including the most influential polynomials in the truncation. Measuring this influence by correlation, the classical greedy algorithms, orthogonal matching pursuit (OMP) \cite{tropp2007signal} and least angle regression (LARS) \cite{efron2004least}, have been utilized to rank the polynomials.   

This contribution is aimed at stabilizing the constructed sparse PCE model with respect to small changes in the training data. Bagging (a.k.a. bootstrap aggregating) \cite{breiman1996bagging} is a popular approach, especially for decision tree methods, to stabilize the modeling approach by training multiple regression models based on bootstrap resamples \cite{efron1994introduction} and taking the final prediction as the mean of all predictions. In the study of variable selection, rather than treating the resamples independently in the construction of regression models, the so-called {\it inclusion frequency} \cite{sauerbrei1992bootstrap} (or {\it inclusion fraction} \cite{royston2009bootstrap}) is computed as the criterion for the importance of a variable. With the knowledge that resamples are perturbed versions of the same original data, the truly important variables should be included in the built model for most bootstrap resamples since all models should reflect the same underlying data structure. The utilization of inclusion frequency improves the replication stability of selected variables \cite{sauerbrei1992bootstrap,AnnaMasterThesis}.

In this paper, the idea of inclusion frequency is applied to the construction of a sparse PCE model. Based on LARS or OMP, multiple PCE models are constructed based on resamples and involved basis polynomials are ranked according to the inclusion frequency. The replication stability of selected polynomials in the final model is expected to be enhanced. Since the PCE model is highly determined by the basis, the stability of the built model would be increased as well. Such construction method of a sparse PCE model is named as {\it resampled PCE} (rPCE).   

Improvements and adjustments are made in rPCE based on the application procedure of inclusion frequency on variable selection. First, recent work in \cite{de2016subsampling} shows subsampling \cite{geisser1975predictive} is superior to bootstrapping in the ability of distinguishing important and redundant variables and in the favor of sparse models. {Remark that subsampling consists of randomly drawing part of samples without replacement while bootstrapping approach generates observations of the same size as the original data but with replacement.} Here, an efficient subsampling technique, $k$-fold division, is applied, where the original data is divided into $k$ parts and a resampling data set is composed of any $k-1$ parts. This procedure ensures the original data is fully explored with $k$ resamples. In variable selection, variables are roughly labeled ``important" or ``redundant"  by comparing the associated inclusion frequency with a cut-off value, the choice of which is still an open problem \cite{walschaerts2012stable}. In rPCE, while the basis polynomials are ranked by inclusion frequency, the number of included polynomials in the final model is decided by cross validation. Moreover, for polynomials with the same inclusion frequency, the associated cross-validation errors are taken as an extra criterion for further ranking. Such ranking approach provides the possibility to combine different basis pursuit methods. Efforts trying to merge the selection results of LARS and OMP are made. 

This paper itself is organized as follows. A general framework of the PCE-based surrogate modeling is introduced in Section \ref{sec:PCE_SurrogateModel}. Section \ref{sec:fPCE&sPCE} gives the concept of the full and sparse PCE truncation, where the building processes based on LARS and OMP are briefly described, respectively. The methodology of rPCE is illustrated in Section \ref{sec:rPCE}. Resampling data through the random division into $k$ parts, based on the generated candidate polynomials by LARS and/or OMP, the importance of polynomials is evaluated through the inclusion frequency. The value of $k$ matters and the determination strategy is discussed in Section \ref{sec:Para}, where the strategy to select the source of candidate polynomials (LARS, OMP, or their combination) is also presented. The improved performances in prediction and sensitivity analysis by rPCE are shown via application to one classical analytical functions, one finite-element model and one finite-difference-time-domain model in Section \ref{sec:examples}. Conclusions and perspectives follow in Section \ref{sec:conclusions}.

\section{Surrogate model based on polynomial chaos expansion}
\label{sec:PCE_SurrogateModel}
\subsection{Probabilistic modeling}
\label{subsec:ProbabilisticModeling}
Consider a physical model represented by a deterministic function $\bm{y}=\mathcal{M}\bracket{\bm{x}}$, where $\bm{x}\in\mathbb{R}^M$ and $\bm{y}\in\mathbb{R}^Q$, $M$, $Q$ being the number of input and output quantities, respectively. The uncertainty of inputs and the propagation to responses lead to the description of $\bm{x}$ and $\bm{y}$ as random vectors, $\bm{X}$ and $\bm{Y}$. Here, since each component of $\bm{Y}$ can be separately analyzed in statistical learning, only cases with scalar response, i.e., $Q=1$, are considered for simplicity.  

Describing the random vector $\bm{X}$ by the joint probability density function (PDF) $p_{\bm{X}}$ and assuming that ${Y}$ has a finite variance, the latter belongs to a Hilbert space $L^2(\mathbb{R}^M,\mathcal{B}_M,\mathbb{P}_{\bm{X}})$, $\mathcal{B}_M$ being the Borel $\sigma$-algebra of the event space $\mathbb{R}^M$ and $\mathbb{P}_{\bm{X}}$ being the probability measure of $\bm{X}$. The Hilbert space is equipped with the following inner product
\begin{equation}
\left\langle f,g\right\rangle =E\sbracket{f(\bm{X})g(\bm{X})}=\int_{\mathbb{X}}f(\bm{x})g(\bm{x})p_{\bm{X}}(\bm{x}) d\bm{x},
\end{equation}
and can be represented by a complete set of orthogonal basis functions.

\subsection{Polynomial chaos expansion}
\label{subsec:PCE}
Polynomial chaos expansion is a spectral representation of $Y$ taking polynomials as basis functions,
\begin{equation}
{Y} = \sum_{\bm{\alpha}\in\mathbb{N}^M}\beta_{\bm{\alpha}}\psi_{\bm{\alpha}}(\bm{X}),
\label{InfinitePCE}
\end{equation} 
where $\bm{\alpha}$ is a vector of non-negative integers indicating the order of multivariate polynomials $\psi_{\bm{\alpha}}$ and $\beta_{\bm{\alpha}}$ is the corresponding expansion coefficient. 

{The construction of $\psi_{\bm{\alpha}}(\bm{X})$ is briefly recalled now \cite{sudret2007uncertainty,soize2004physical}}. Assuming that the input random variables are independent, the multivariate polynomials is a tensor product of univariate polynomials $\pi_{\alpha_i}$, i.e.,
\begin{equation}
\psi_{\bm{\alpha}}(\bm{X}) = \pi_{{\alpha}_1}^{(1)}(X_1) \times \ldots \times \pi_{{\alpha}_M}^{(M)}(X_M),
\end{equation}
where {$\pi_{\alpha_i}^{(i)}$'s are univariate orthonormal polynomials with respect to the PDF of the $i$-th parameter $X_i$, with degree $\alpha_i$ (e.g., Hermite polynomials for Gaussian distributions). This methodology is referred to as generalized PCE (gPCE) \cite{soize2004physical,xiu2002wiener}.} For PDFs not included in gPCE, a nonlinear mapping of input variables to the known ones can be made with the technique of isoprobabilistic transformation \cite{lebrun2009generalization,lemaire2013structural} or specific orthogonal polynomials are computed numerically via the Stieltjes procedure \cite{gautschi2004orthogonal}.  

The PCE coefficients $\beta_{\bm{\alpha}}$ {are} obtained in a non-intrusive way by the regression approach. A data set $\{\bm{x}^{(n)},n=1,\ldots,N\}$ sampled from the input PDF $p_{\bm{X}}$ and the corresponding response $\{y^{(n)}=\mathcal{M}(\bm{x}^{(n)})\}$ compose altogether the ED. With notations of column vector $\bm{y}=[y^{(n)}]$, $\bm{\beta}=[\beta_{\bm{\alpha}}]$ and matrix $\bm{\psi}=[\psi_{\bm{\alpha}}(\bm{x}^{(n)})]$, the PCE coefficients can be obtained from
\begin{equation}
\hat{\bm{\beta}} = \arg \min_{\bm{\beta}} ||\bm{y}-\bm{\psi}\bm{\beta}||_2^2,
\end{equation}
which yields the ordinary least square (OLS) \cite{rao1973linear} solution as { the normal equation}
\begin{equation}
\hat{\bm{\beta}} = \bracket{\bm{\psi}^T\bm{\psi}}^{-1}\bm{\psi}^T\bm{y},
\label{OLS}
\end{equation}
the superscript ``$T$" denoting the transpose operation. {Remark that overfitting problems may be suffered for the OLS solution but can be avoided by implementing regularization \cite{jakeman2015enhancing,tibshirani1996regression} techniques, as provided by the LARS algorithm described later on.}

Remark that, although only cases with independent inputs are considered in the above analysis, it is possible to describe the mutual dependence by a copula \cite{nelsen2007introduction} and use Rosenblatt transformation \cite{lebrun2009generalization} to cast the problem as a function of auxiliary independent variables.

\subsection{Estimation of prediction performance}
\label{subsec:predictionPerformance}
{The model assessment is often performed by Monte Carlo simulations with a large {test dataset}, which is independent from the experimental design. Denote $\widehat{\mathcal{M}}$ as the surrogate model, the input vector and response of the $n$-th {test} data as $\bm{x}_{\text{test}}^{(n)}$ and $y_{\text{test}}^{(n)}$, respectively. The performance of the constructed model is assessed by computing the mean square error of data discrepancy} 
\begin{equation}
\epsilon_{\text{test}} = \frac{1}{N_{\text{test}}}\sum_{n=1}^{N_{\text{test}}}\bracket{\mathcal{M}(\bm{x}_{\text{test}}^{(n)})-\widehat{\mathcal{M}}(\bm{x}_{\text{test}}^{(n)})}^2.
\end{equation} 
For an easier interpretation of $\epsilon_{\text{test}}$, the associated coefficient of determination $R^2_{\text{test}}$ is computed by
\begin{equation}
R^2_{\text{test}} = 1-\frac{\epsilon_{\text{test}}}{\mathrm{Var}(\bm{y}_{\text{test}})},
\label{def_R2}
\end{equation}
where $\mathrm{Var}(\bm{y}_{\text{test}})=\sum_{n=1}^{N_{\text{test}}}(y_{\text{test}}^{(n)}-\bar{y}_{\text{test}})^2/(N_{\text{test}}-1)$ and $\bar{y}_{\text{test}}=\sum_{n=1}^{N_{\text{test}}}y_{\text{test}}^{(n)}/N_{\text{test}}$. Therefore, the closer $R^2_{\text{test}}$ is to one, the more accurate is the prediction by $\widehat{\mathcal{M}}$.

However, in scenarios with high computational cost for a single simulation, it is usually intractable to have a large {test dataset}. Then, the same data as for training are often reused for {model assessment}. However, the underestimation of the generalization error is well-known in the case of overfitting \cite{blatman2009adaptive}. {Cross-validation} was thus proposed and is commonly advocated \cite{kohavi1995study,konakli2016polynomial}. {Here, leave-one-out {cross-validation (LOOCV)} is applied and the corresponding cross-validation error reads:} 
\begin{equation}
\epsilon_{\text{LOO}} = \frac{1}{N}\sum_{n=1}^{N}\bracket{\mathcal{M}(\bm{x}^{(n)})-\widehat{\mathcal{M}}^{-(n)}(\bm{x}^{(n)})}^2,
\label{LOO}
\end{equation}
where $\widehat{\mathcal{M}}^{-(n)}$ denotes the surrogate model trained by leaving the $n$-th data out. Remark that $\epsilon_{\text{LOO}}$ is also known as predicted residual of squares (PRESS) or jacknife error \cite{efron1982jackknife} and it can be computed fast in single training process \cite{blatman2009adaptive} by
\begin{equation}
\epsilon_{\text{LOO}} = \frac{1}{N}\sum_{n=1}^{N}\bracket{\frac{\mathcal{M}(\bm{x}^{(n)})-\widehat{\mathcal{M}}(\bm{x}^{(n)})}{1-h_n}}^2,
\label{fastLOO}
\end{equation}
where $h_n$ is the $n$-th diagonal element of the matrix $\bm{\psi}\bracket{\bm{\psi}^T\bm{\psi}}^{-1}\bm{\psi}^T$.

\section{Surrogate modeling based on full PCE and sparse PCE}
\label{sec:fPCE&sPCE}
The accurate PCE of the true model is an infinite series and needs a truncation for the sake of computation. From Eq. \eqref{InfinitePCE}, one sees that truncating a PCE is actually selecting a subset of $\mathbb{N}^M$ for $\bm{\alpha}$ such that the system response can be represented by the associated polynomials at a sufficient accuracy. Assuming the selected $\bm{\alpha}$ vectors compose the set $\mathbb{A}$, the truncated PCE can be written as
\begin{equation}
\widehat{\mathcal{M}}(\bm{X}) =  \sum_{\bm{\alpha}\in\mathbb{A}}\beta_{\bm{\alpha}}\psi_{\bm{\alpha}}(\bm{X}).
\label{TruncatePCE}
\end{equation} 
{Setting a maximum value to the total degree of polynomials} leads to the so-called full PCE model, which suffers from the curse-of-dimensionality \cite{friedman2001elements}, meaning that the cardinality of $\mathbb{A}$ sharply increases with the number of input parameters, as explained below. {While the problem of curse-of-dimensionality can be moderated by the algorithm of Smolyak sparse quadrature \cite{smolyak1963quadrature},} recently least angle regression (LARS) \cite{blatman2011adaptive,marelli2015uqlab} and orthogonal matching pursuit (OMP) \cite{tropp2007signal,marelli2015uqlab} have been used to downsize the truncation and achieve the so-called sparse PCE model. 

\subsection{Full PCE model}
\label{subsec:fPCE}
$\mathbb{A}$ is commonly selected by setting a maximum to the total degree of multivariate polynomials, i.e., $\mathbb{A}_{full}=\{\bm{\alpha}\in\mathbb{N}^M, \sum_{i=1}^{M}\alpha_i \le  p\}$, $p$ a positive integer. The PCE-based surrogate model with this setup is named in the sequel as the full PCE model. However, the cardinality of $\mathbb{A}_{full}$, denoted by $P_{full}$, equals $\binom{p+M}{p}$ and polynomially increases with the value of $p$ and $M$. Moreover, to ensure the well-conditioning of the information matrix $\bm{\psi}$ in Eq.~\eqref{OLS}, the ED size $N$ should be larger than $P_{full}$. As a result, the resulting curse of dimensionality prevents the application of the full PCE model in scenarios with large $p$ and $M$. This problem is addressed by downsizing $\mathbb{A}$ through the use of greedy algorithms, so that only the most influential polynomials are included in the truncated PCE. 

\subsection{Sparse PCE model}

{The problem of curse-of-dimensionality  during the construction of full PCE models is addressed by constructing the so-called sparse PCE models, where $\mathbb{A}$ is downsized through the use of greedy algorithms, so that only the most influential polynomials are included in the truncated PCE.  
	
	\begin{table}[!ht]
		\centering
		\caption {Procedures of constructing a sparse PCE model, $J_{\text{max}}=\min\{N-1,\text{card}(\mathbb{A}_{\text{full}})\}$.}
		\begin{tabular}{c l}
			\toprule
			&For $p=1,\ldots,p_{\text{max}}$,\\
			&\hspace{7pt}1. $\mathbb{A}_{\text{full}}=\{\bm{\alpha}\in\mathbb{N}^M, \sum_{i=1}^{M}\alpha_i \le  p\}$ and set active set $\mathbb{A}_0^a=\varnothing$;\\[4pt]
			&\hspace{7pt}2. Rank basis polynomials in $\{\psi_{\bm{\alpha}},\bm{\alpha}\in\mathbb{A}_{\text{full}}^p\}$ by LARS, OMP, or rPCE. The $\bm{\alpha}$ corresponding with\\
			&\hspace{7pt}the first $J_{\text{max}}$ most influential basis polynomials compose the set 
			$\{\bm{\alpha}_j,j=1,\ldots,J_{\text{max}}\}$.\\[4pt]
			&\hspace{7pt}3. For $j=1,\ldots,J_{\text{max}}$, \\
			&\hspace{25pt}Update $\mathbb{A}_j^a=\mathbb{A}_{j-1}^a\cup\bm{\alpha}_j$. Based on $\bm{\psi}_{\mathbb{A}_j^a}$, compute $\bm{\beta}_j$ as the OLS solution and associated $\epsilon_{\text{LOO}}^{j}$.\\
			&\hspace{19pt}End\\
			&\hspace{7pt}4. ${J}=\arg\min_j \{\epsilon_{\text{LOO}}^{j}\}$ and $\epsilon_{\text{LOO}}^{p,\text{min}}=\epsilon_{\text{LOO}}^{J}$. When $p\ge 3$, if $\epsilon_{\text{LOO}}^{p,\text{min}}>\epsilon_{\text{LOO}}^{p-1,\text{min}}>\epsilon_{\text{LOO}}^{p-2,\text{min}}$, stop the\\
			&\hspace{7pt}model-construction process and output the PCE model corresponding with $\bm{\psi}_{\mathbb{A}_{J}^a}$.\\
			&End\\
			\bottomrule
		\end{tabular}
		\label{Algorithm_sparsePCE}
	\end{table}
	
	Table~\ref{Algorithm_sparsePCE} presents the procedures to construct a sparse PCE model. Based on candidate $\bm{\alpha}$ from the full PCE model, i.e., $\bm{\alpha}\in\mathbb{A}_{\text{full}}$, the associated basis polynomials are ranked (e.g., by correlation with response data $\bm{y}$ for OMP in Table~\ref{Algorithm_OMP}) and the first $J_{\text{max}}$ most influential ones are selected by OMP (refer to algorithm in Table~\ref{Algorithm_OMP}) or LARS (refer to Table~\ref{Algorithm_LARS}), $J_{\text{max}}$ being the maximum number of included polynomials in the final constructed PCE model and set as $\min\{N-1,\text{card}(\mathbb{A}_{\text{full}})\}$ (otherwise the least-square problem becomes ill-posed).
	
	Assessing the model performance by leave-one-out cross-validation, the optimal number of selected polynomials, $J$, corresponds with the PCE model with the minimal  $\epsilon_{\text{LOO}}$, the computation of which follows \eqref{fastLOO}, where only the surrogate model constructed with the whole set of data is required.
	
	The optimal value for the total degree of polynomials follows an early-stopping criterion. Setting the maximum value for $p$, a progressive increase stops when the minimal $\epsilon_{\text{LOO}}$ increase with two consequent $p$.}

\section{Surrogate modeling based on resampled PCE}
\label{sec:rPCE}
During replications with resampled training data, different PCE truncations are obtained by LARS or OMP and the inclusion frequency of involved polynomials can be computed. Resampled PCE (rPCE) is proposed to refine standard PCE truncation schemes by making use of the inclusion frequency. Cross-validation error associated with each polynomial is an additional factor to further rank polynomials with the same inclusion frequency. Efforts to combine selection results by LARS and OMP to further improve the performance of rPCE are also presented.

\subsection{Resampled PCE based on LARS or OMP}
\label{subsec:rPCE_or}

Inclusion frequency is defined as the percentage of replications \cite{sauerbrei1992bootstrap} in which a given basis polynomial is selected by LARS or OMP. {The variation of training data is simulated by the subsampling technique, $k$-fold division,} considering its efficiency in exploiting the information of original data, i.e., the training process makes use of all data in $k$ replications.

Dividing the whole set of data into $k$ subsets, all with approximately same size. Of $k$ subsets, the $l$-th subset is left out and the remaining $k-1$ subsets are used for the PCE construction. Varying $l$ from $1$ to $k$, one {has} $k$ PCE models built  by LARS/OMP and the associated active sets are denoted by $\mathbb{A}_{P,(l)}^{a}$, $l=1,\ldots,k$. The subscript ``$P$" and superscript ``$a$" are ignored in $\mathbb{A}_{P,(l)}^{a}$ to be $\mathbb{A}_{(l)}$ in the followings.

To search for the most frequent $\bm{\alpha}$ indices within the $k$ different sets $\mathbb{A}_{(l)}$, $l=1,\ldots,k$, one can merge the latter into a {multiset} $\mathbb{A}^{\text{Mul}} = \{\mathbb{A}_{(1)},\ldots,\mathbb{A}_{(k)}\}$, {the superscript ``Mul" denoting a multiset (rather than set), which allows for multiple instances for each $\bm{\alpha}$.} Then the selection frequency of $\bm{\alpha}$ in the $k$ building processes is equal to the number of its duplicates in $\mathbb{A}^{\text{Mul}}$.  {Denote $\mathbb{A}$ as the set (thus no duplicate elements) composed of elements in  $\mathbb{A}^{\text{Mul}}$. The selection frequency corresponding with each of element in $\mathbb{A}$ is an integer in the interval $[1,k]$ and saved in the vector $\bm{s}_f$}. The inclusion frequency is computed as the normalized frequency, i.e., $\bm{s}_f/k$.

For applications wherein the idea of inclusion frequency has been applied, the final model keeps components ({e.g., influential variables for the variable-selection problem \cite{sauerbrei1992bootstrap}}) for which the inclusion frequency exceeds the cutpoint $\nu$. The value of $\nu$ impacts much on the stability and complexity of the final model but is usually arbitrarily taken \cite{de2016subsampling,walschaerts2012stable}, and no conclusive method seems available for an optimal choice of $\nu$ \cite{sauerbrei1992bootstrap,AnnaMasterThesis}. To avoid this problem, based on the ranked polynomials, the total number of included polynomials {(rather than the cutpoint $\nu$)} in the final model is chosen by cross validation {following the procedures in Table~\ref{Algorithm_sparsePCE}.}  

However, during the running of rPCE, different {multi-indices} $\bm{\alpha}$ might have the same frequency, which introduces some uncertainty in the ranking of polynomials. To avoid this uncertainty, one more factor, namely the effect of each basis polynomial on $\epsilon_{\text{LOO}}$, is considered. 

From the LARS/OMP procedures, one can see that the correlated polynomials are sequentially added into the active set, thus the increment of $\epsilon_{\text{LOO}}$ by adding $\bm{\alpha}_j$ into $\mathbb{A}^a_{j-1}$ equals $\Delta \epsilon_{\text{LOO}}^{j} = \epsilon_{\text{LOO}}^{j}-\epsilon_{\text{LOO}}^{j-1}$ for $j\ge 1$, where $\epsilon_{\text{LOO}}^{0}$ is set as $0$. {Thus, each $\bm{\alpha}$ in $\mathbb{A}^{\text{Mul}}$ corresponds with a  $\Delta \epsilon_{\text{LOO}}$.}

{Add the superscript ``($l$)" to the notation standing for the quantity obtained by leaving the $l$-th subset out from model construction.}
Then, the so-called {\it error score} $\mathbf{s}_e$ can be computed as the mean of all terms $\Delta \epsilon_{\text{LOO}}^{(l),j}$ mapping to the same element of $\mathbb{A}$, i.e.,
\begin{equation}
s_e^i=\frac{1}{s_f^i\Delta \epsilon_{\text{LOO}}^{\max}}\sum_{\{{(l),j}| \bm{\alpha}^{(l),j}=\bm{\alpha}^i\}} \Delta \epsilon_{\text{LOO}}^{(l),j}, \,  i=1,\ldots,\text{card}\{\mathbb{A}\}.
\end{equation}
where the superscript ``$i$" stands for the $i$-th element of a vector or set. The normalization by $\Delta \epsilon_{\text{LOO}}^{\max}$, the maximum element of {$|\Delta \epsilon_{\text{LOO}}^{(l),j}|$}, is to confine the value of $s_e^i$ between $-1$ and $1$ such that the ranking of polynomials by the total score 
\begin{equation}
\mathbf{s}=\mathbf{s}_f + \mathbf{s}_e,
\end{equation}
is mainly affected by $\mathbf{s}_f$ in rPCE. Remark that $\mathbf{s}_f$ is used instead of inclusion frequency (the normalized $\mathbf{s}_f$) and is {subsequently named} {\it frequency score}.

\subsection{Resampled PCE combining LARS and OMP}
\label{subsec:rPCE_and}
The way to rank polynomials in rPCE allows the possibility to combine the results by LARS and OMP. Following the procedures in Section \ref{subsec:rPCE_or}, $\mathbb{A}^{\text{Mul}}$ and $\mathbb{E}^{\text{Mul}}$ {(multiset of $\Delta \epsilon_{\text{LOO}}^{(l),j}$)} can be obtained by LARS and OMP separately, denoted by $\mathbb{A}^{\text{Mul,LARS}}$, $\mathbb{E}^{\text{Mul,LARS}}$ and $\mathbb{A}^{\text{Mul,OMP}}$, $\mathbb{E}^{\text{Mul,OMP}}$, respectively. Then, merging results by LARS and OMP into a single {multiset}, $\mathbb{A}^{\text{Mul}}=\{\mathbb{A}^{\text{Mul,LARS}},\mathbb{A}^{\text{Mul,OMP}}\}$ and $\mathbb{E}^{\text{Mul}}=\{\mathbb{E}^{\text{Mul,LARS}},\mathbb{E}^{\text{Mul,OMP}}\}$, from which $\mathbb{A}$ and the associated total score $\bm{s}$ can be computed. Then, {the basis polynomials associated with $\mathbb{A}$ are ranked according to $\bm{s}$ and the construction of a sparse PCE model follows procedures in Table~\ref{Algorithm_sparsePCE}.} 

\section{Parameter settings}
\label{sec:Para}
\subsection{Resampling scheme}
\label{subsec:resampling}
The $k$-fold division is used to simulate the data variation in rPCE and the value of $k$ matters on the performance.
A tradeoff lies behind the determination of $k$. With a small $k$ (e.g., $k=2$), a large portion (half) of data is apart from the building process. As a result, some information of the true system might be lost or not accurately learned by the surrogate model and the selected polynomials may not be truly influential. On the other side, a large $k$, (e.g., $k=N$) cannot sufficiently simulate the data statistical variation and the selected polynomials in the construction of the $k$ different PCEs might have a high correlation. This way, the polynomials selected by rPCE would be almost the same as those with LARS or OMP and the prior knowledge, from which rPCE is to benefit, cannot be well exploited.   

The proposed strategy is to merge $\mathbb{A}^{\text{Mul}}$ obtained for different values of $k$. Considering that the validation error on the data left out is used to estimate the prediction performance in Section \ref{subsec:polynomialSource} and values of $3,5,10,20, N$ (leave-one-out), are usually recommended \cite{kohavi1995study,anderssen2006reducing,fushiki2011estimation} for $k$-fold cross validation, rPCE will run based on the {multiset} $\mathbb{A}^{\text{Mul}} = [\mathbb{A}_3^{\text{Mul}},\mathbb{A}_5^{\text{Mul}},\mathbb{A}_{10}^{\text{Mul}},\mathbb{A}_{20}^{\text{Mul}}, \mathbb{A}_N^{\text{Mul}}]$, where {the subscript of $\mathbb{A}_q^{\text{Mul}}$ corresponds with the value of $k$}. Data variation is fully simulated via $k=3,5$ and the bias error is small considering that in average about $0.86N$ resamples (without replacement) are used to generate candidate polynomials. It seems not easy to optimize the setting of $k$, especially considering that the optimal value may differ w.r.t. scenarios. However, the proposed setting is revealed robust in the various application examples.

{With respect to a set of $k$ values, i.e., $k=\{3,5,10,20,N\}$, the total score can be computed based on $\bm{s}_{f,k}$ and $\bm{s}_{e,k}$, the subscript ``$k$" indicating the quantity for a specific value of $k$.} Denote  $\mathbb{A}$ as the copy of $\mathbb{A}^{\text{Mul}}$ but without element duplication. For each $\bm{\alpha}$ in $\mathbb{A}$, its selection frequency can be computed by 
\begin{equation}
{f^i} = \sum_{k=\{3,5,10,20,N\}}s^i_{f,k}, \, i=1,\ldots,\text{card}(\mathbb{A}),
\label{frequencyScore}
\end{equation} 	
where the superscript ``$i$" stands for the $i$-th element of a vector and $s^i_{f,k}$ equals zero if the $i$-th $\bm{\alpha}$ of $\mathbb{A}$ is not in $\mathbb{A}_k^{\text{Mul}}$. Since $s^i_{f,k}$ is upper bounded by $k$, the polynomials selected with small values of $k$ (e.g., elements in $\mathbb{A}_3^{\text{Mul}}$) will have small {values of $f^i$} and be less likely to have high ranks in rPCE.

To solve this problem, instead of \eqref{frequencyScore}, the frequency score is computed as a  summation of weighted $s^i_{f,k}$:
\begin{equation}
s_f^i = \sum_{k=\{3,5,10,20,N\}}s^i_{f,k}\frac{\text{lcm}(3,20,N)}{k}, \, i=1,\ldots,\text{card}(\mathbb{A}),
\label{weightedFrequencyScore}
\end{equation} 
where $\text{lcm}(3,20,N)$ computes the least common multiple of $3, 20, N$ (same for $3,5,10,20,N$). The weights give rise to the same maximum value of the summands in \eqref{weightedFrequencyScore}. Consequently, the candidate polynomials w.r.t. different values of $k$ are equally considered in rPCE.

Finally, the set of $k$ values, i.e., $\{3,5,10,20,N\}$, needs an adjustment for a small $N$. For instance, $k$ can only be $3,5,10,N$ when $N=15$. 

{The computation of error score follows as:
	\begin{equation}
	s_e^i=\frac{1}{f^i}\sum_{k=\{3,5,10,20,N\}} s_{e,k}^i, \,  i=1,\ldots,\text{card}\{\mathbb{A}\},
	\end{equation}
	where $s^i_{e,k}$ equals zero if the $i$-th $\bm{\alpha}$ of $\mathbb{A}$ is not in $\mathbb{A}_k^{\text{Mul}}$.
} 

\subsection{Source of candidate polynomials}
\label{subsec:polynomialSource}

Section \ref{sec:rPCE} presents the rPCE based on candidate polynomials generated by three sources, LARS, OMP or their combination, and one needs to decide which source is the optimal option. The polynomials commonly and frequently selected by two different approaches are believed influential and more likely to be included in rPCE. However, if one approach has a much worse performance than the other, the combination scheme would not be recommended, since the candidate polynomials generated by the worse approach might deteriorate the performance of rPCE.  Therefore, if LARS is much better than OMP, only candidate polynomials by LARS participate into the ranking in rPCE, and vice versa. Otherwise, the combination scheme is used. 

The criterion of ``much better" should be properly set. Assuming a large set of validation data is available, as illustrated in Section \ref{subsec:predictionPerformance}, $R^2_{\text{test}}$ can be computed as the unbiased estimation of the prediction performance. Here, the comparison of two building approaches is conducted with the analysis of the distribution of $R^2_{\text{test}}$. Varying the training data, a sequence of surrogate models is built and the associated $R^2_{\text{test}}$ values are computed. Representing $\mathbb{R}^2_{\text{test,LARS}}$ and $\mathbb{R}^2_{\text{test,OMP}}$ as the sets of $R^2_{\text{test}}$ values obtained by LARS and OMP respectively, the first and third quartile of these two sets are computed and denoted by $Q_1^{\text{LARS}}$, $Q_1^{\text{OMP}}$, $Q_3^{\text{LARS}}$, $Q_3^{\text{OMP}}$. Then, if $Q_1^{\text{LARS}}>Q_3^{\text{OMP}}$, one considers that LARS is much better than OMP, and vice versa. Otherwise, LARS and OMP are considered with similar performances and the combination scheme would be adopted.

However, again a large set of validation data is usually not available due to the high computational costs. Here, $R^2_{\text{test}}$ is approximated through the validation on the data left out in the $k$-fold division. With different values of $k$ and $l$, the validations generate a set of  determination coefficient ${R}^2_{k,(l)}$ as the approximations to $R^2_{\text{test}}$, $l=1,\ldots,k$, $k\in\{3,5,10,20,N\}$. Denoting 
${\mathbb{R}}_{\text{LARS}}^2$ and ${\mathbb{R}}_{\text{OMP}}^2$ as
the sets of ${R}^2_{k,(l)}$  values obtained by LARS and OMP, 
the distribution of sets $\mathbb{R}^2_{\text{test}}$ is then simulated by ${\mathbb{R}}_{\text{LARS}}^2$ and ${\mathbb{R}}_{\text{OMP}}^2$.
\begin{figure*}[!h]
	\centering
	\includegraphics[width=0.9\linewidth]{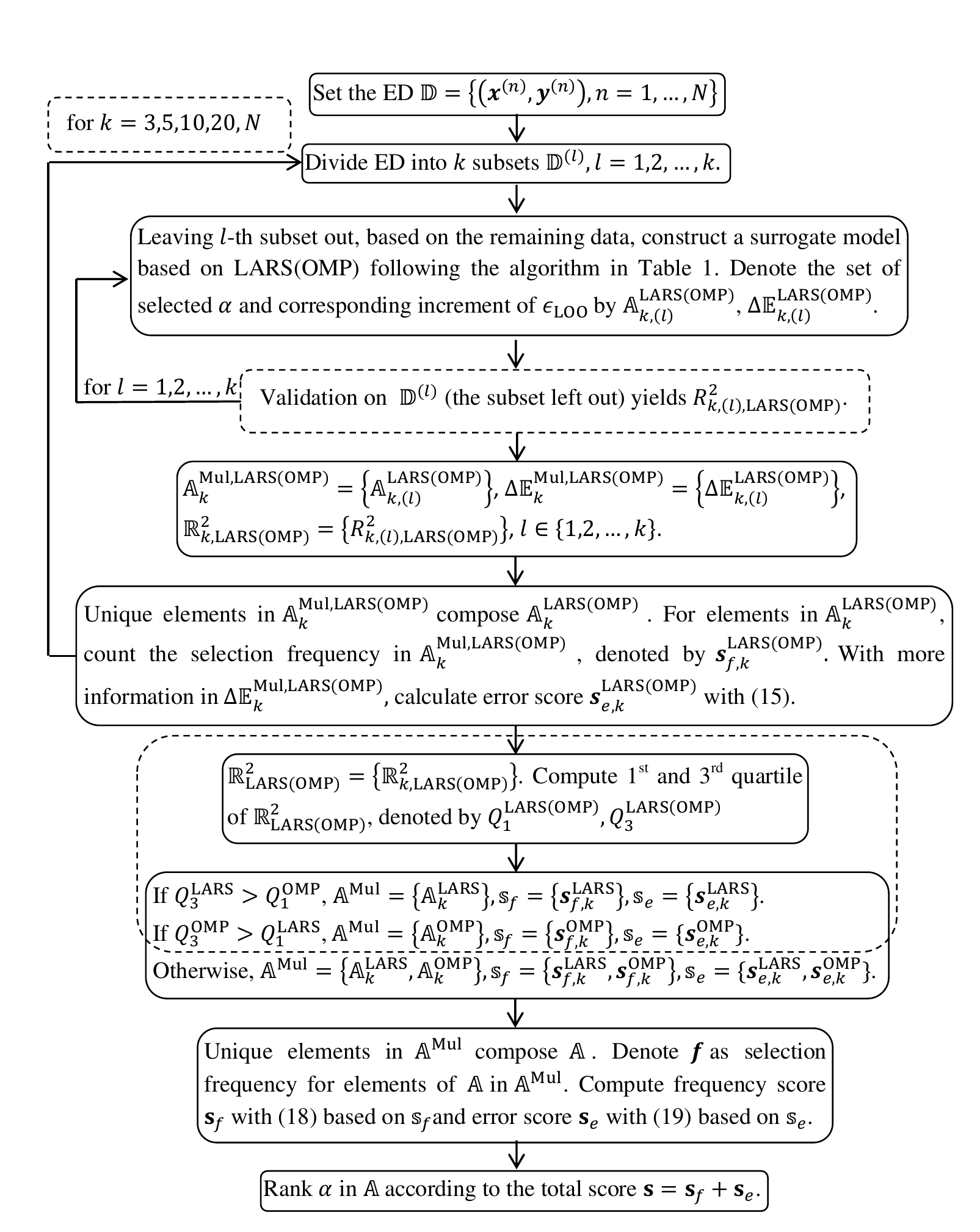}
	\caption{ Flow chart for ranking basis polynomials based on resampled PCE, where steps enclosed by dashed lines are with the suggested configurations in Section \ref{sec:Para}.}
	\label{flowChart}
\end{figure*}
Remark that two layers of cross validations now have been operated in rPCE. The outer cross validation is just illustrated to simulate the distribution of $\mathbb{R}_{\text{LARS}}^2$ and $\mathbb{R}_{\text{OMP}}^2$. The inner one is embedded in the running of LARS and OMP to compute $\epsilon_{\text{LOO}}$ in Table~\ref{Algorithm_OMP} and \ref{Algorithm_LARS}. The two-layer cross validation here is indeed an realization of the known {\it double-cross-validation} (DCV) \cite{baumann2014reliable} or {\it cross model validation} (CMV) \cite{anderssen2006reducing,gidskehaug2008cross}. The related literature shows the unbiased estimation of $R^2_{\text{test}}$ by the determination coefficient from the outer cross-validation errors, i.e., ${R}^2_{k,(l)}$.

The procedures to {rank basis polynomials} by rPCE are summarized in Fig. \ref{flowChart}. {Then, the construction of sparse PCE models follows the steps in Table~\ref{Algorithm_sparsePCE}.}

Benefiting from the obtained PCE model, the global sensitivity analysis, which measures the impacts of input variables to the response, can be conducted via the computation of Sobol' indices \cite{sobol1993sensitivity,homma1996importance} for independent variables or Kucherenko indices \cite{kucherenko2012estimation} for dependent cases by Monte-Carlo simulations. Note that in the case of independent inputs, Sobol' indices are readily available from PCE coefficients, as shown in \cite{sudret2008global}.

\section{Application examples}
\label{sec:examples}
The knowledge that the influential polynomials are to be frequently selected during replications is first checked on
a specially designed function, the true basis polynomials of which are known. Then, to present the performance of surrogate modeling based on rPCE and the comparisons to LARS and OMP, two benchmark functions (with dimension $M=3$ and $M=8$, respectively), a finite-element model (with $M=10$) and a finite-difference-time-domain model (with $M=4$) are analyzed. The PCE models based on LARS and OMP are obtained with the Matlab package UQLab {(www.uqlab.com)} \cite{marelli2014uqlab,webUQLab}, {where the maximum degree of multivariate polynomials $p$ is set as $20$}. Using resampling, UQLab provides the candidate polynomials to rPCE. { Remark that if no specific configurations are given in the following examples, resampled PCE is performed with the suggested configurations in Section \ref{sec:Para}, i.e., optimized source (LARS, OMP, or both) of candidate polynomials and candidate polynomials from  $k=\{3,5,10,20,N\}$.}

Latin-Hypercube sampling \cite{mckay1979comparison} is used to sample the input random variables. Since cases with a small ED are concerned in this paper, the size of ED $N$ is chosen between $10$ and $50$ here. As mentioned in Section \ref{subsec:PCE}, dependent variables can be analyzed after the transformation into the corresponding independent ones through the generalized Nataf transformation, so only examples with independent variables are presented in this section and the global sensitivity is analyzed with the computation of Sobol' indices.

\subsection{Summation of multivariate polynomials}
\label{subsec:multiPoly}
\begin{figure}[!ht]
	\centering
	\subfigure[]{\includegraphics[width = 0.5\linewidth]{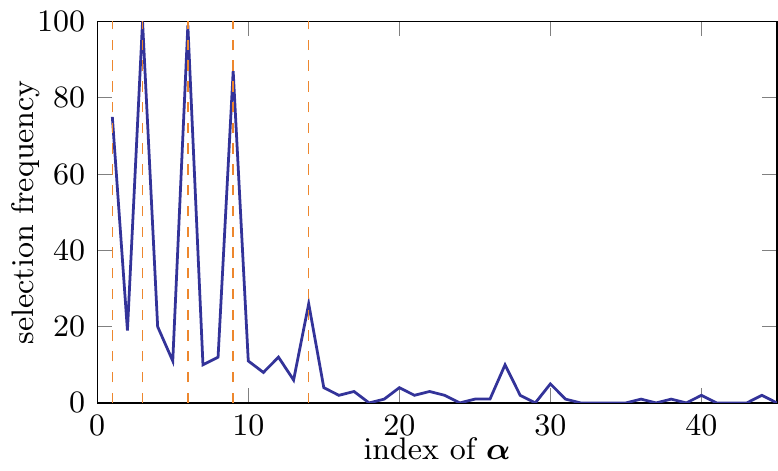}\label{linearPoly_freq}}~
	\subfigure[]{\includegraphics[width = 0.5\linewidth]{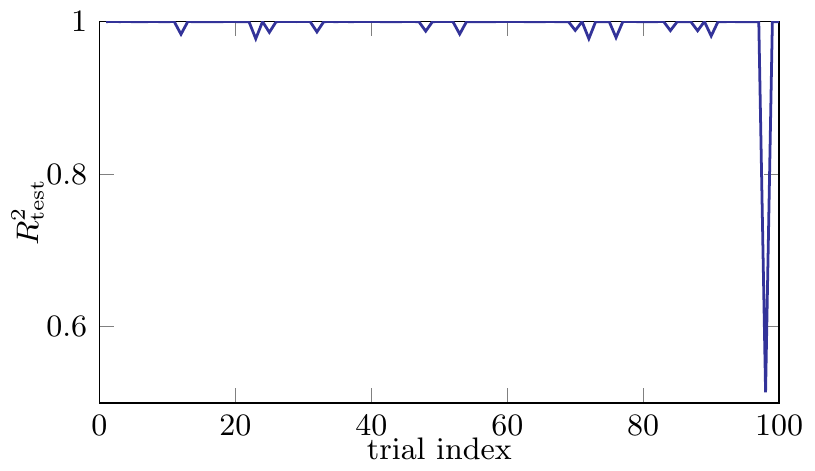}\label{linearPoly_R2}}
	\caption{Example 1: Summation of multivariate polynomials - (a) the selection frequency of $\bm{\alpha}$ by OMP and (b) the associated $R^2_{\text{test}}$ in all replications}
	\label{linearPoly}
\end{figure}

To show that the influential polynomials associated with the true model are frequently selected, the surrogate modeling of the following expression, 
\begin{equation}
Y = 1 + X_1 + X_1 X_2 + X_1 X_2^2 + X_1 X_2^3,
\label{multiPoly}
\end{equation}
which is a summation of five multivariate polynomials (including the constant term), is conducted. $X_1$ and $X_2$ are independent variables that follow the Gaussian distributions $\mathcal{N}(0,1)$ and $\mathcal{N}(6,1)$, respectively. OMP is used to build a sparse PCE model with $12$ data points for training and $10^4$ data for independent {testing}. A total of $100$ PCE constructions are made to check the selection frequency of polynomials.

Due to the Gaussian distribution of input variables, Hermite polynomials are used to compose the basis, where the bivariate polynomials are indexed by $\bm{\alpha}=(\alpha_1,\alpha_2)$. 
The constant term corresponds with $\bm{\alpha}=(0,0)$, while the other four terms in Eq. \eqref{multiPoly} are with $(1,0), (1,1),(1,2),(1,3)$, respectively. Labeling $\bm{\alpha}$ by integers, the selection frequency during the $100$ PCE constructions is plotted in Fig. \ref{linearPoly_freq}, where the dashed lines indicate the five true $\bm{\alpha}$ indices. Remark that, the selection frequency is smaller than $2$ when the labels are larger than $45$ and only the results with labels $\le45$ are displayed for a better visualization. As observed, although the true indices of $\bm{\alpha}$ are not always selected, they are the most frequent ones during replications. Making use of this knowledge and selecting the most frequent $\bm{\alpha}$ (also the associated polynomial) may improve the performance of the obtained PCE model and avoid the outliers (for example the $98$-th replication with ${R}_{\text{test}}^2 = 0.51$ in Fig.~\ref{linearPoly_R2}, where $X_2$, $X_1$, $X_1^3$ are selected as the basis).   

\subsection{Ishigami function}
\label{subsec:Ishigami}

The Ishigami function, which is defined by
\begin{equation}
Y = \sin X_1 + a \sin^2 X_2 + bX_3^4\sin X_1,
\label{defIshigami}
\end{equation} 
is widely used for benchmarking in uncertainty and sensitivity analysis. The parameters are set to $a=7$, $b=0.1$ and the input random variables $X_i$, $i = 1,2,3$, are independent and uniformly distributed over $[-\pi, \pi]$. Legendre polynomials are thus used as the basis according to the principle of the generalized PCE. 

\begin{figure}[!ht]
	\centering
	\subfigure[rPCE ($R_{\text{test}}^2=0.9971$)]{\includegraphics[width = 0.33\linewidth]{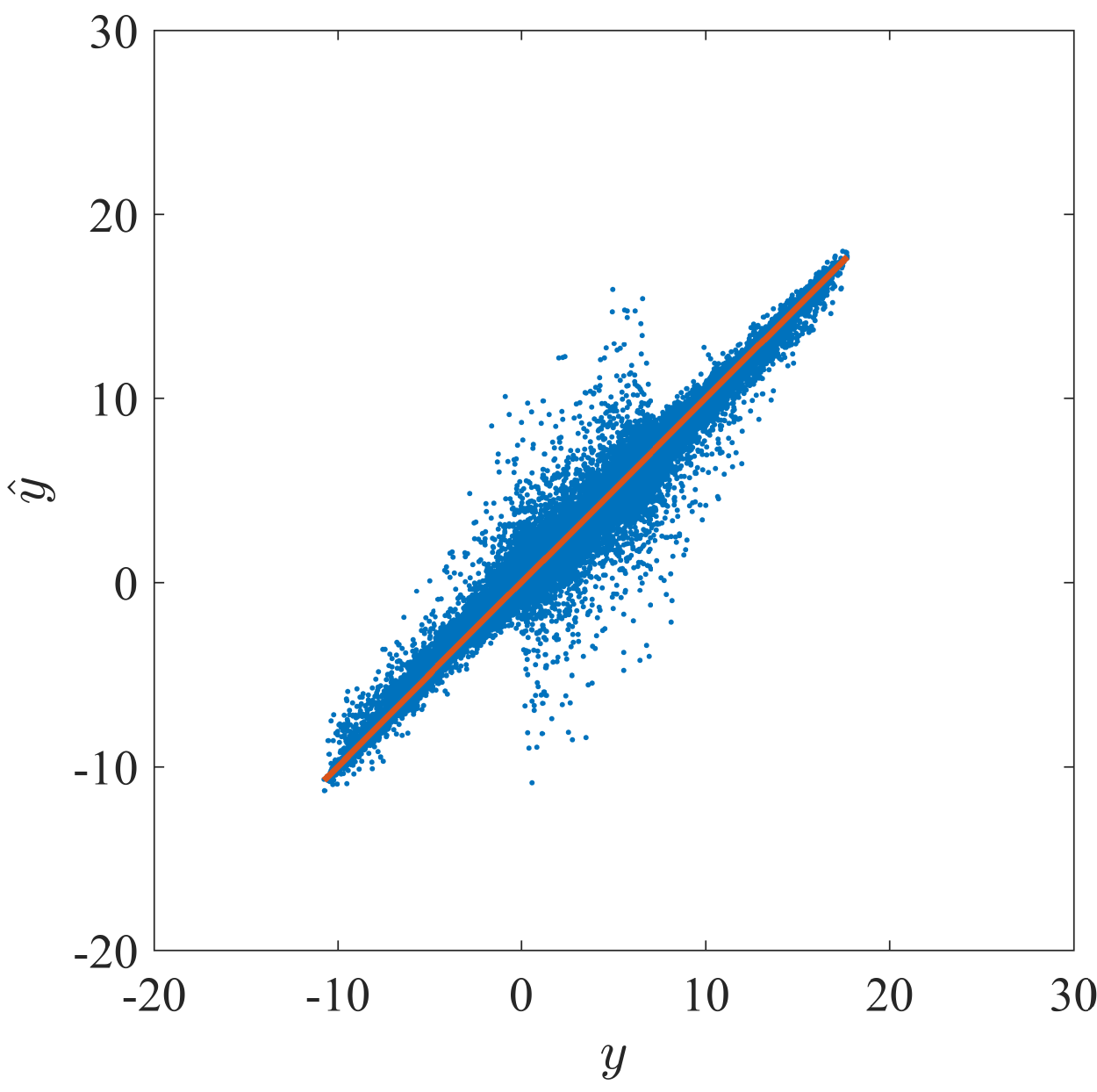}\label{diagIshigami_rPCE}}~
	\subfigure[LARS ($R_{\text{test}}^2=0.8724$)]{\includegraphics[width = 0.33\linewidth]{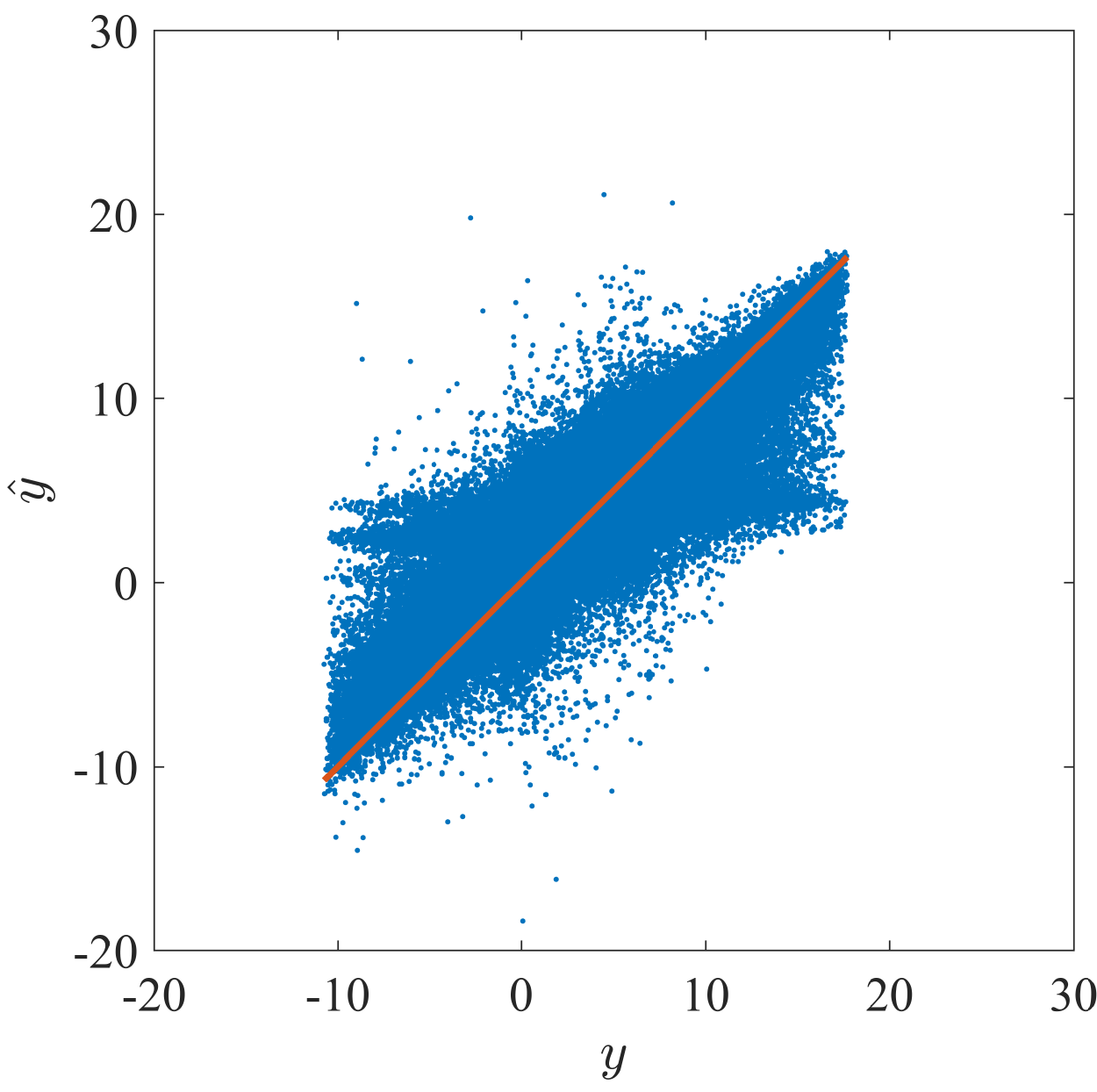}\label{diagIshigami_LARS}}~
	\subfigure[OMP ($R_{\text{test}}^2=0.8790$)]{\includegraphics[width = 0.33\linewidth]{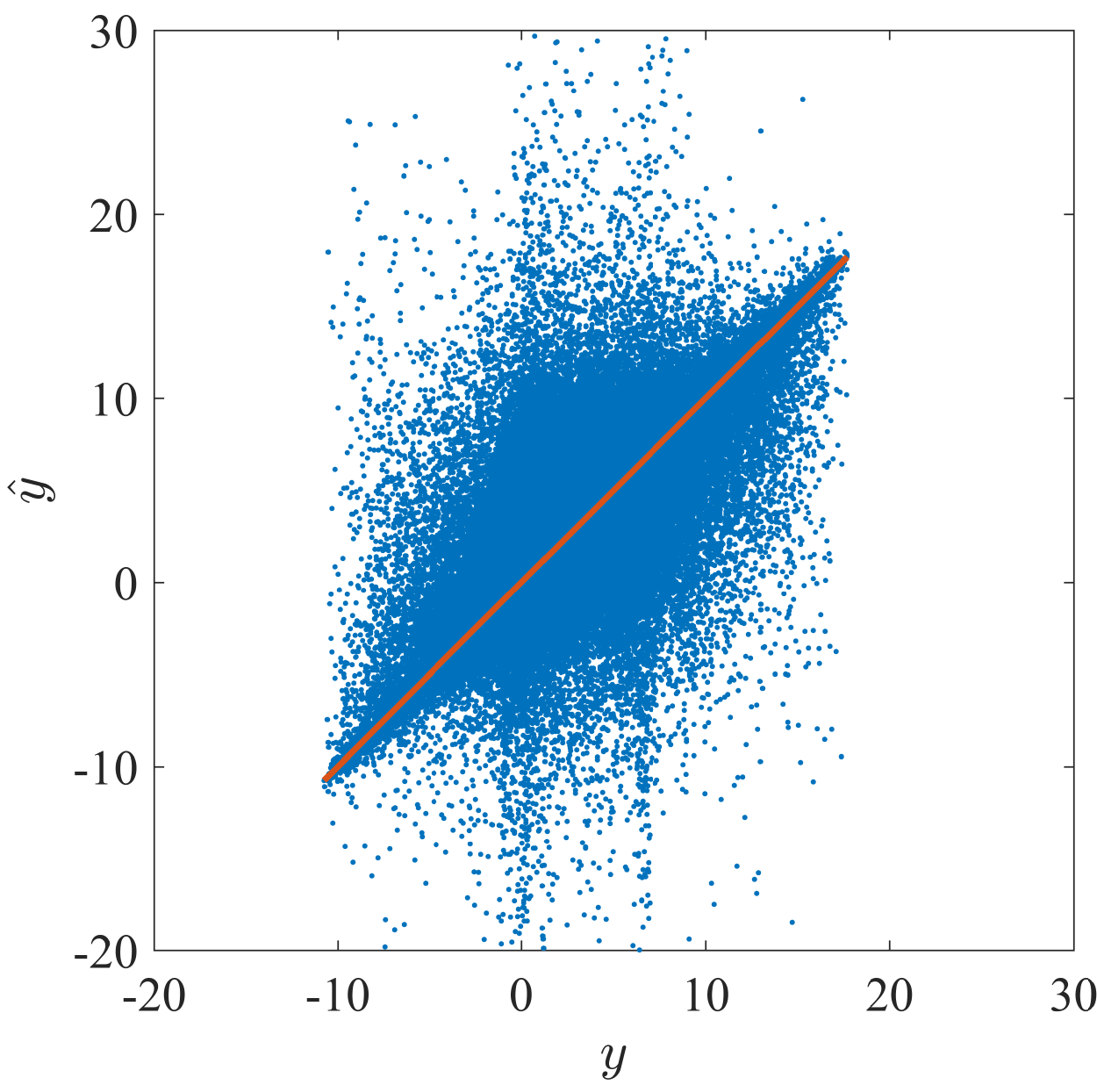}\label{diagIshigami_OMP}}
	\caption{Ishigami function - prediction of validation data by (a) rPCE, (b) LARS and (c) OMP with $50$ data points ($100$ replications).}
	\label{diagIshigmai}
\end{figure}

First, $50$ data points are used for building the surrogate model and $10^4$ points for estimating the prediction performance. The analysis is repeated $100$ times in order to investigate the statistical uncertainty of different modeling approaches. The prediction of all validation data ($10^6$ data over $100$ replications) by the surrogate models built based on LARS, OMP and rPCE is shown in Fig.~\ref{diagIshigmai}, where $y$ stands for the true value, $\hat{y}$ for the predicted one, and the solid line indicates the case when $\hat{y}$ exactly equals $y$. As observed, although rPCE and OMP provide unbiased estimations of the Ishigami function, OMP suffers from more outliers and a higher variance. LARS tends to have larger predictions (relative to the true values) when $y<0$ and smaller predictions when $y>8$. Meanwhile, the prediction variance of LARS is not as small as rPCE. 

\begin{figure}[!ht]
	\centering
	\includegraphics[width =0.55\linewidth]{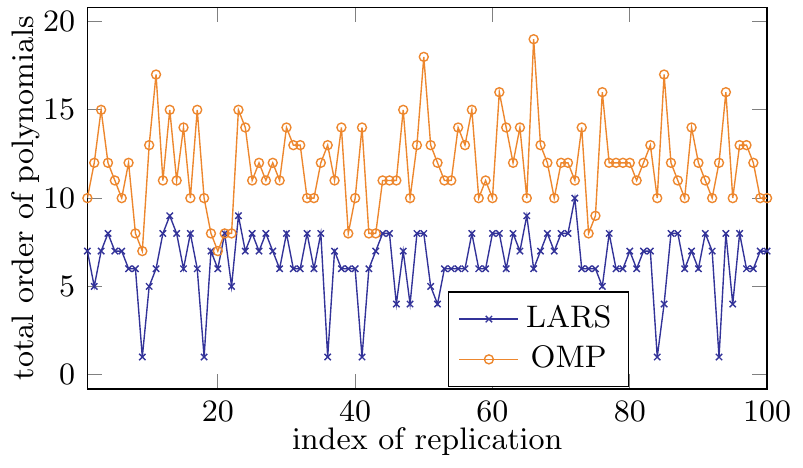}
	\caption{Ishigami function - optimal total order of polynomials selected by LARS and OMP in $100$ replications.}
	\label{Ishigmai_totalOrder}
\end{figure}	
{The reason for different performances of surrogate modeling based on LARS and OMP may be seen from Fig.~\ref{Ishigmai_totalOrder}. The PCE models are with higher orders when constructed based on OMP than based on LARS. The larger value of the total order $p$ leads to a bigger polynomial basis and thus a more flexible surrogate model, which tends to have less biased but high-variance predictions.}

\begin{figure}[!ht]
	\centering
	\includegraphics[width =0.9\linewidth]{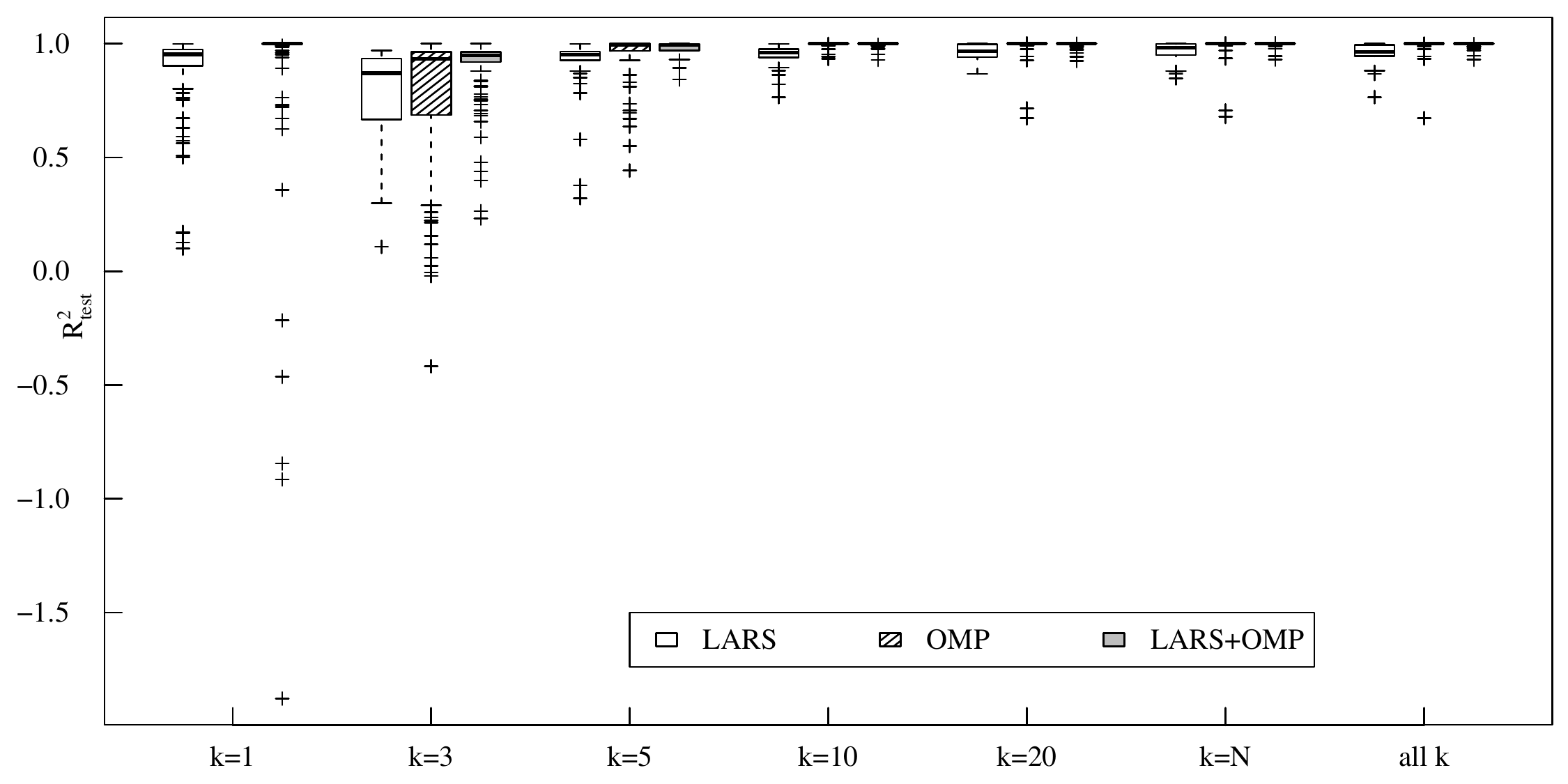}
	\caption{Ishigami function - box plots of $R_{\text{val}}^2$ using different values of $k$ in $k$-fold division with $50$ data points ($100$ replications).}
	\label{Ishigmai_boxplot}
\end{figure}
As mentioned in Section \ref{sec:Para}, statistical uncertainty is emulated via the $k$-fold division in rPCE and the value of $k$ matters. The suggested configuration of rPCE is combining the polynomial-selection results with $k=\{3,5,10,20,N\}$. To show the effects of $k$, $R^2_{\text{test}}$ is computed at each replication and $100$ values of $R^2_{\text{test}}$ yield the box plots of Fig.~\ref{Ishigmai_boxplot}, where $k=1$ indicates the surrogate modeling with the whole set of training data but without the refinement by rPCE and ``all $k$" denotes the rPCE results by combining results with different values of $k$. As observed, when $k=1$, although the interquartile range (IQR), i.e., the span between the first quartile to the third quartile, of LARS is larger than that of OMP, more outliers appear with OMP and the minimum $R^2_{\text{test}}$ is even smaller than $-1.5$. With rPCE, except the case of $k=3$, improvements can be observed from the reduced outliers and/or prediction variance. The combination of LARS and OMP, denoted by ``LARS+OMP" (see Section \ref{subsec:rPCE_and}), seems to have advantages over the rPCE based on LARS or OMP and the advantages are more obvious with cases $k=3$ and $5$.    

\begin{table}[!ht]
	\centering
	\caption{Ishigami function - mean of $R_{\text{val}}^2$ over $100$ replications with $50$ data points ($100$ replications).}
	\begin{tabular}{l|*{3}{c}}
		\toprule
		&{LARS}&{OMP}&{LARS+OMP}\\
		\midrule
		$k=1$ & $0.8723$ &0.8788&\\
		$k=3$ & 0.7890 & 0.7734 & 0.8935 \\
		$k=5$ & 0.9281 & 0.9566 & 0.9817 \\
		$k=10$ & 0.9542 & 0.9972 & 0.9974 \\
		$k=20$ & 0.9630 & 0.9919 & 0.9969 \\
		$k=N$ & 0.9686 & 0.9918 & 0.9978 \\
		all $k$ & 0.9619 & 0.9947 & 0.9971\\
		\bottomrule
	\end{tabular}
	\label{Ishigami_table}
\end{table}
As quantitative comparisons, Table \ref{Ishigami_table} gives the mean of $R_{\text{test}}^2$ over $100$ replications. Generally, OMP is better than LARS. However, the advantage of OMP is not large and, as a result, the combination of LARS and OMP in rPCE generates better surrogate models. Remark that the means in Table \ref{Ishigami_table} are obtained by fixing the value of $k$ and the source of candidate polynomials (LARS, OMP, or LARS+OMP) during all replications. Selecting the ``all $k$" option and optimizing the polynomial source at each replication with the suggested configuration in Section \ref{sec:Para}, the obtained mean of $R_{\text{test}}^2$ equals $0.9972$, only $6\times10^{-4}$ smaller than the highest value when $k=N$ with LARS+OMP. 

\begin{figure}[!ht]
	\centering
	\includegraphics{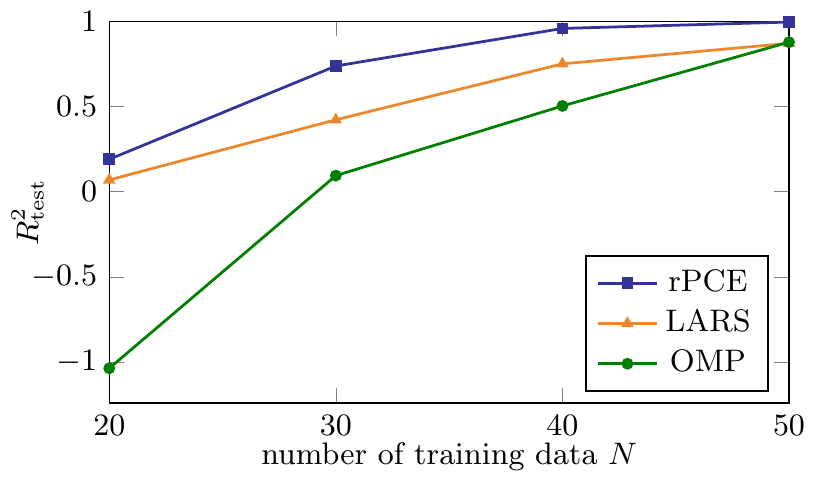}
	\caption{Ishigami function - mean of $R_{\text{val}}^2$ versus different values of $N$ ($100$ replications).}
	\label{IshigamiR2_N}
\end{figure}
Simulations with $N=20, 30, 40$ are also operated with the same configurations and the means of $R_{\text{test}}^2$ are plotted as the line graph in Fig.~\ref{IshigamiR2_N}, which shows the better performance of rPCE compared to LARS and OMP in the cases with small EDs.  
\begin{figure*}[!ht]
	\centering
	\includegraphics[width =0.85\linewidth]{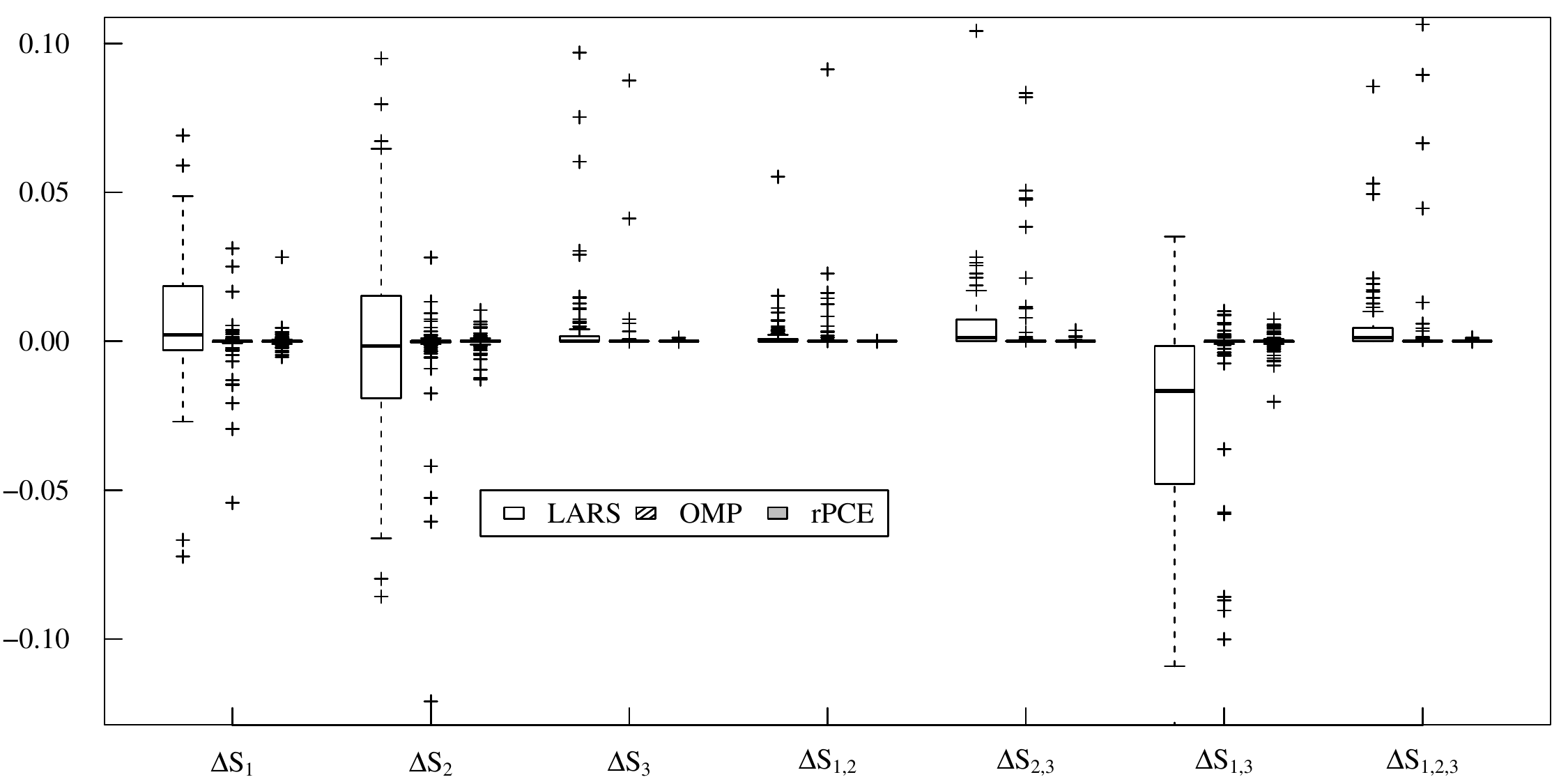}
	\caption{Ishigami function - the estimation error of Sobol' indices with $50$ data points ($100$ replications).}
	\label{SU_Ishigami_N50}
\end{figure*}

The Sobol' sensitivity indices can be analytically computed according to
\begin{equation}
\begin{aligned}
&D = \frac{a^2}{8}+\frac{b\pi^4}{5}+\frac{b^2\pi^8}{18}+\frac{1}{2},\,\\
&D_1 = \frac{b\pi^4}{5}+\frac{b^2\pi^8}{50}+\frac{1}{2},\,\\&D_2 = \frac{a^2}{8},\, D_{1,3}=\frac{8b^2\pi^8}{225},\\
&D_3=D_{1,2}=D_{2,3}=D_{1,2,3}=0. 
\end{aligned}
\end{equation}
Taking the analytical solution as the reference, the estimation error of the Sobol' indices by the PCE-based surrogate model is computed by 
\begin{equation}
\Delta S_i = S_i^{\text{PCE}} - S_i^{\text{ref}},
\end{equation}
where the superscripts of $S$ indicate the generation approach. With $N=50$ and $100$ replications, the box plots of all $\Delta S_i$ are shown in Fig.~\ref{SU_Ishigami_N50}, { where only values between $-0.12$ and $0.1$ are presented for a better view and several outliers are absent.} The variance of $\Delta S_i$ is relatively large with LARS when the Sobol' indices are non zero, i.e., $\Delta S_1$, $\Delta S_2$, $\Delta S_{1,3}$, and the outliers are efficiently avoided by rPCE. The mean of $S_i$ is given by Table~\ref{table_SU_Ishigami_N50}, from which the superiority of rPCE in the sensitivity analysis of the Ishigami function is obviously observed. The accuracy of rPCE for estimating Sobol' indices is in the order of $10^{-4}$ when using $50$ data points in the experimental design.  
\begin{table}[!t]
	\centering
	\caption{Ishigami function - mean of Sobol' indices $50$ data points ($100$ replications).}
	\begin{tabular}{l|*{4}{c}}
		\toprule
		&Reference&{rPCE}&{LARS}&{OMP}\\
		\midrule
		$S_1$ & 0.3139 & 0.3141 & 0.3553 & 0.3017\\
		$S_2$ & 0.4424 & 0.4422 & 0.4152 & 0.4239\\
		$S_3$ & 0.0000 & 0.0000 & 0.0114 & 0.0028\\
		$S_{1,2}$ & 0.0000 & 0.0000 & 0.0017 & 0.0052\\
		$S_{2,3}$ & 0.0000 & 0.0001 & 0.0096 & 0.0042\\
		$S_{1,3}$ & 0.2437 & 0.2435 & 0.2019 & 0.2363\\
		$S_{1,2,3}$ & 0.0000 & 0.0001 & 0.0049 & 0.0258\\
		\bottomrule
	\end{tabular}
	\label{table_SU_Ishigami_N50}
\end{table}

\subsection{Maximum deflection of a truss structure}
\label{subsec:truss}

\begin{figure}[!ht]
	\centering
	\includegraphics[width=0.5\linewidth]{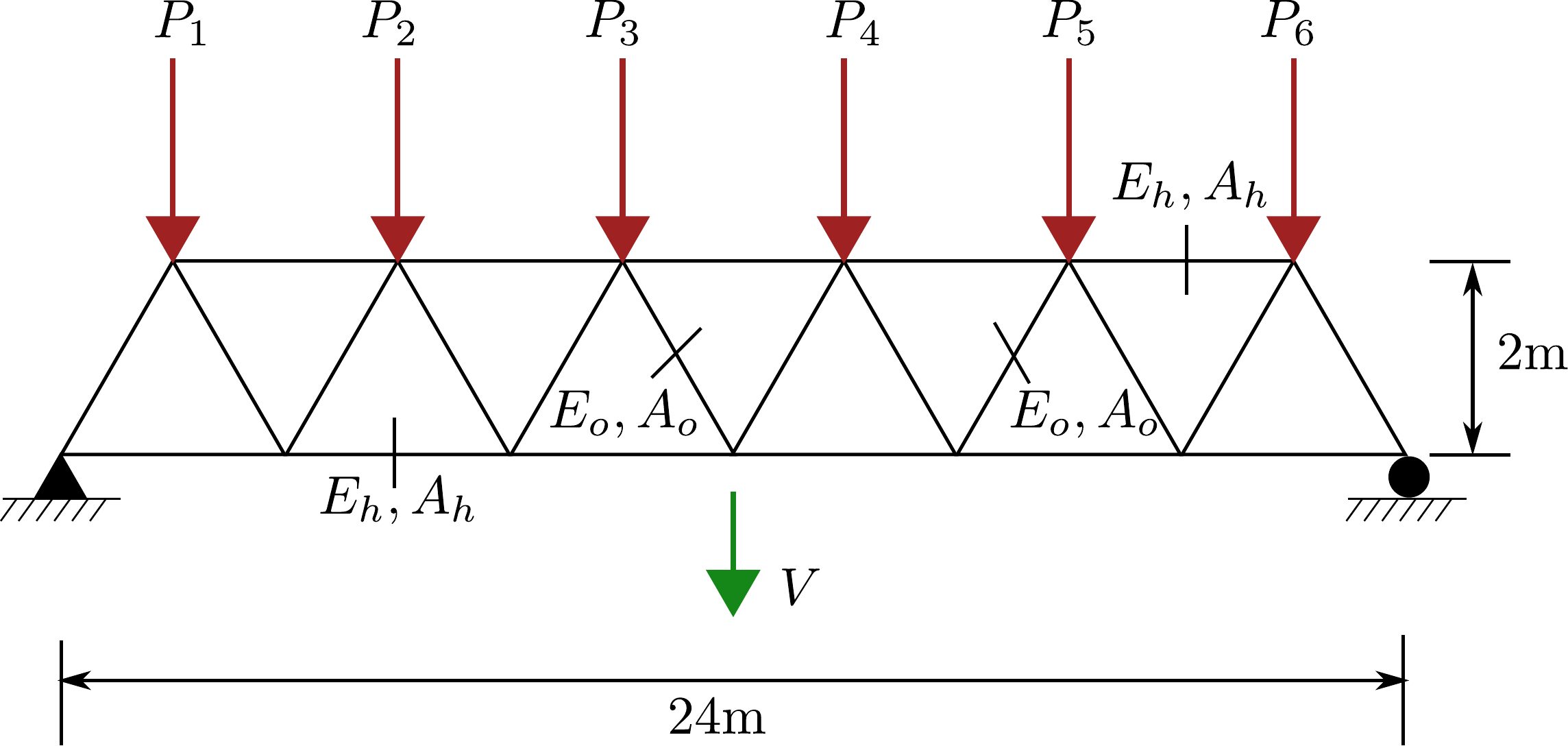}
	\caption{Sketch of a truss structure made of 23 bars \cite{blatman2011adaptive}.}
	\label{truss_structure}
\end{figure}
In Fig.~\ref{truss_structure}, six vertical loads denoted by $P_1\sim P_6$ are put on a truss structure composed of $23$ bars, the cross-sectional area and Young's modulus of which are respectively denoted by $A$ and $E$, the subscripts ``$h$" and ``$o$" standing for the horizontal and oblique bars. The response quantity of interest, the mid-span deflection $V$, is computed with the finite-element method. 

\begin{table*}[!ht]
	\centering
	\caption{Truss deflection - description and distribution of input variables \cite{blatman2011adaptive}.}\label{Truss_pdf}
	\begin{tabular}{lllll}
		\toprule
		Variable & {Distribution}&{Mean}& Std & Description\\
		\midrule
		$E_h,E_o$ (Pa) & Lognormal & $2.1\times 10^{11}$ & $2.1\times 10^{10}$& Young's moduli\\
		$A_h$ (m$^2$) & Lognormal & $2.0\times 10^{-3}$ & $2.0\times 10^{-4}$& cross-section area of horizontal bars\\
		$A_o$ (m$^2$) & Lognormal & $1.0\times 10^{-3}$ & $1.0\times 10^{-4}$& cross-section area of oblique bars\\
		$P_1 \sim P_6$ (N)& Gumbel & $5.0\times 10^4$ & $7.5\times 10^{3}$& vertical loads\\
		\bottomrule
	\end{tabular}
\end{table*}

To analyze the uncertainty of the response, the input parameters are modeled by ten independent random variables following the distributions in Table~\ref{Truss_pdf}. Transforming the input variables into standard normal ones with the isoprobabilistic transformation, LARS, OMP and rPCE surrogate models are built with basis composed of Hermite polynomials. 

\begin{figure}[!ht]
	\centering
	\subfigure[rPCE ($R_{\text{test}}^2=0.9770$)]{\includegraphics[width = 0.33\linewidth]{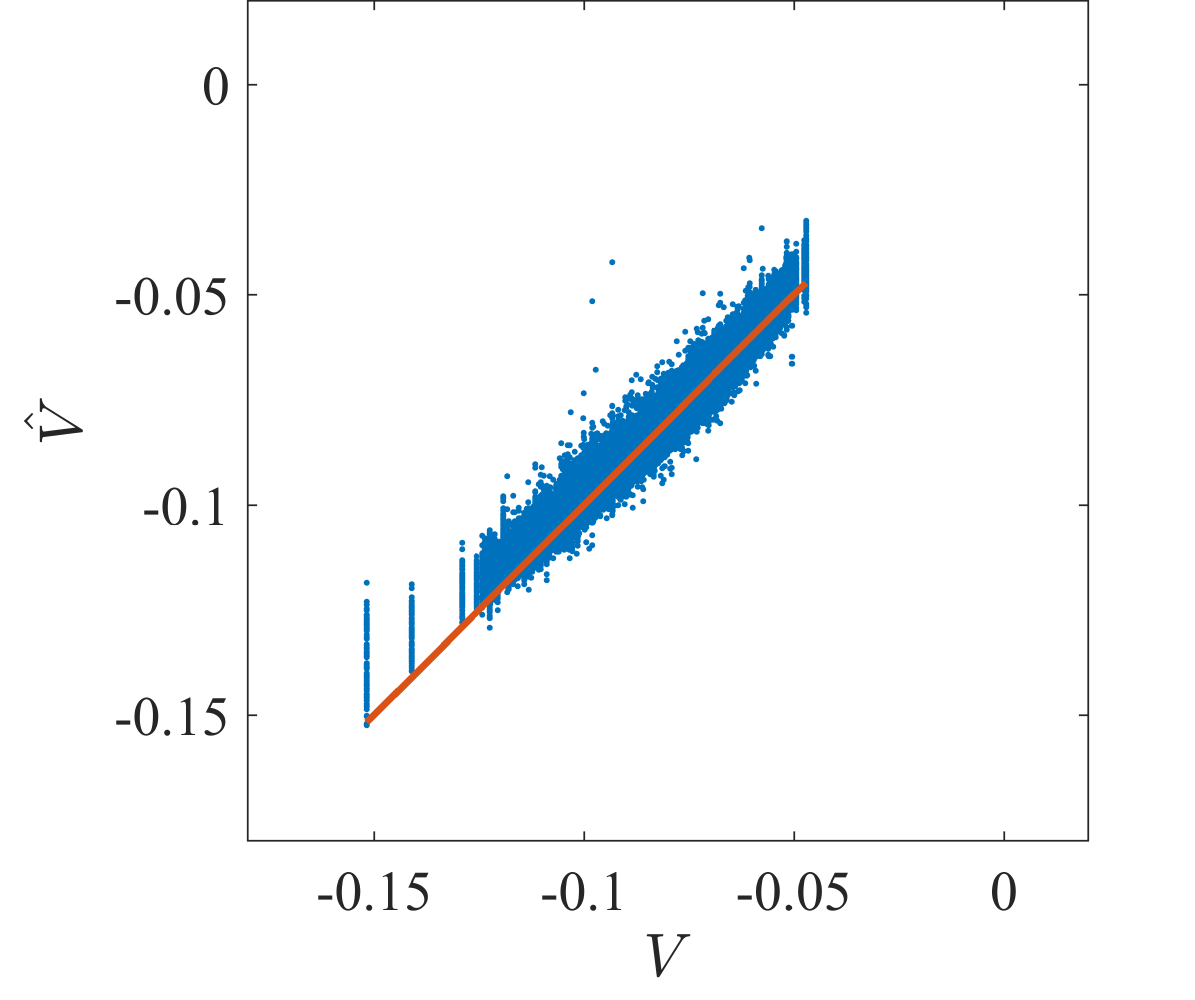}\label{diagTruss_rPCE}}~
	\subfigure[LARS ($R_{\text{test}}^2=0.9631$)]{\includegraphics[width = 0.33\linewidth]{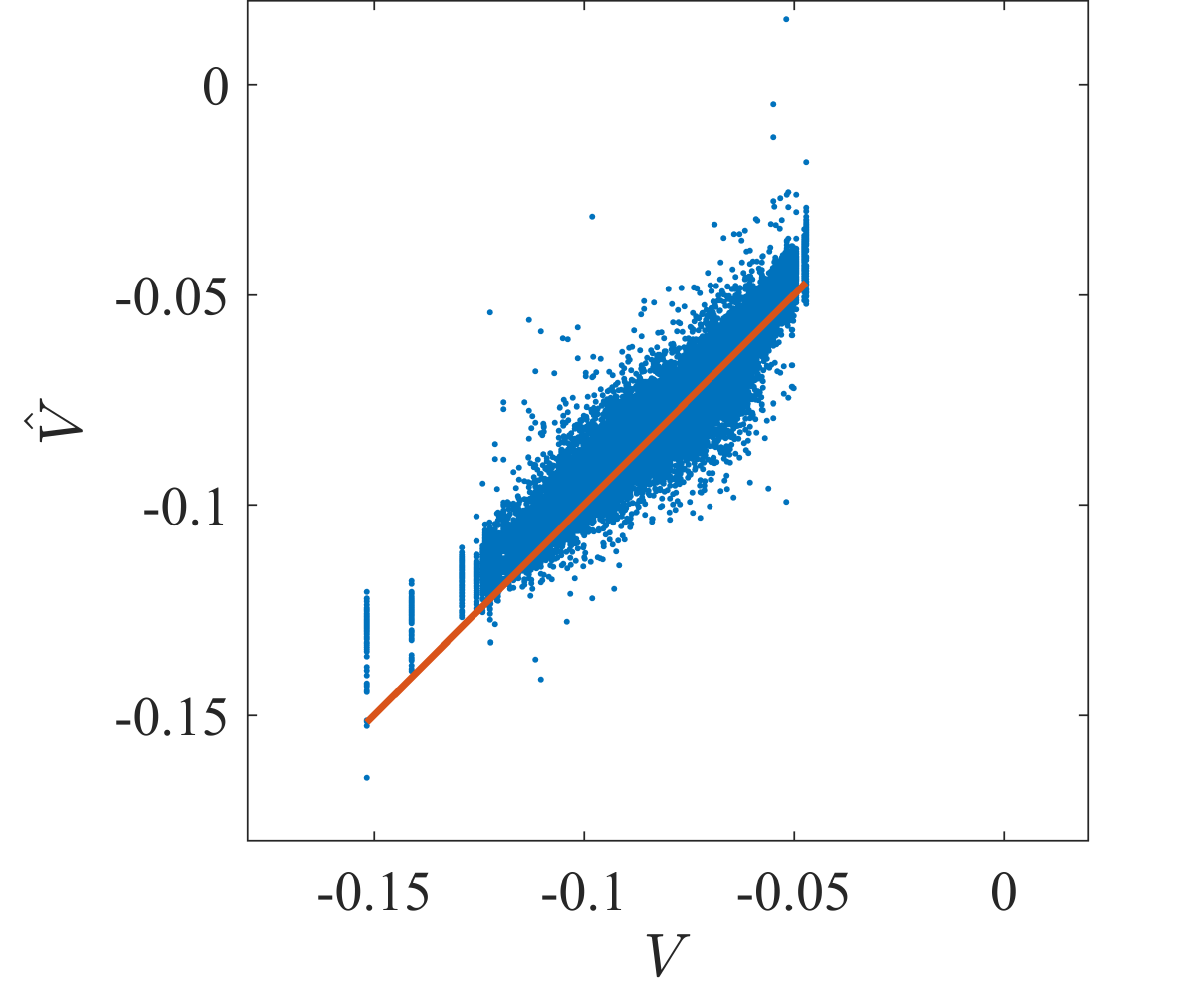}\label{diagTruss_LARS}}~
	\subfigure[OMP ($R_{\text{test}}^2=-6.2257$)]{\includegraphics[width = 0.33\linewidth]{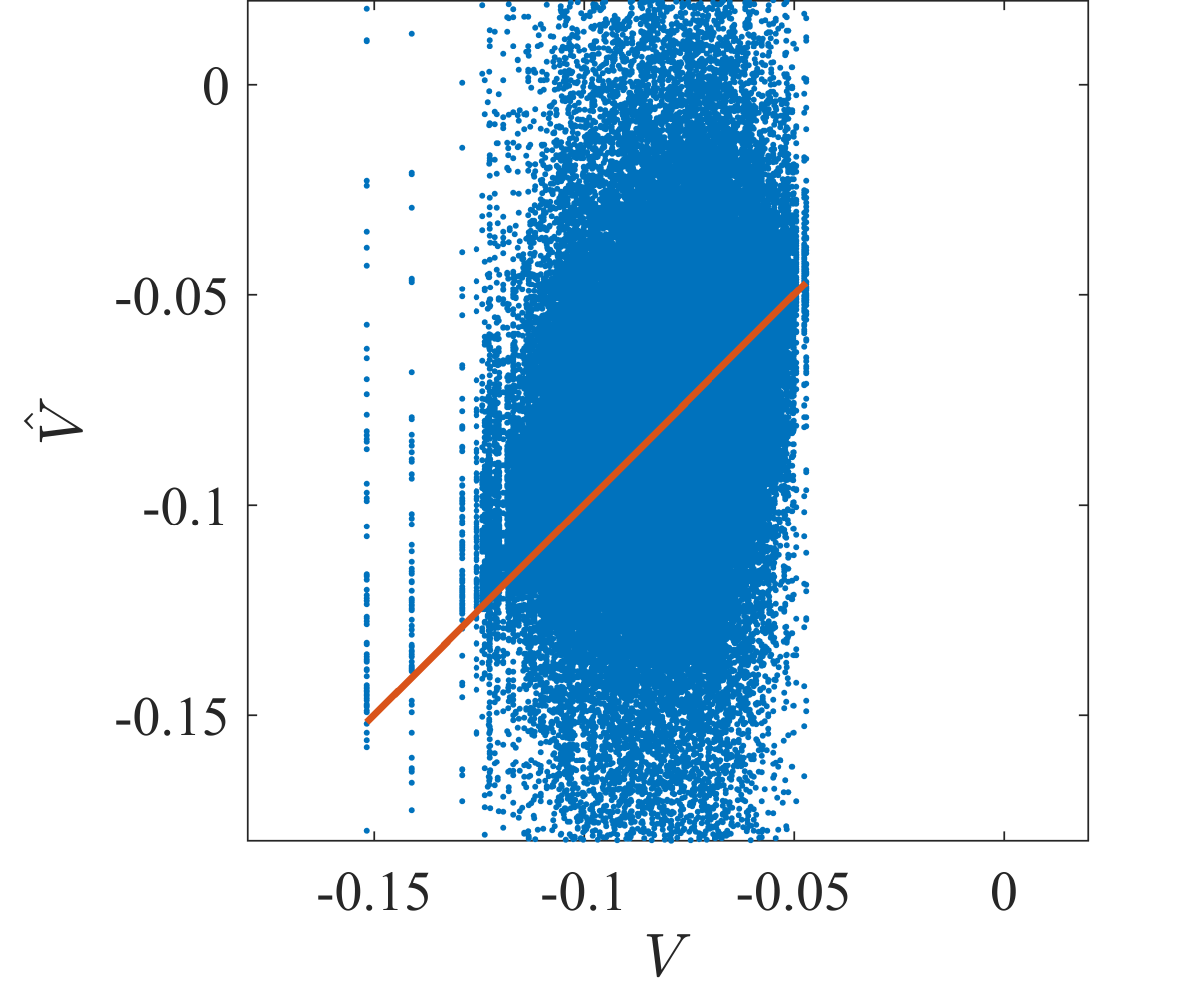}\label{diagTruss_OMP}}
	\caption{Truss deflection - prediction of validation data by (a) rPCE, (b) LARS and (c) OMP with $50$ data points ($100$ replications).}
	\label{diagTruss}
\end{figure}

With $N=50$ and $10^4$ data for validation at each replication, Fig.~\ref{diagTruss} shows the prediction results by the surrogate models over $100$ replications and the solid line indicates the true values of $V$. OMP definitely fails in this scenario. Although the predictions are unbiased, the variance is high due to the too much flexibility of the PCE model built by OMP. In contrast, LARS and rPCE achieve a much better trade-off between the variance and bias. Moreover, rPCE is slightly superior to LARS in variance and the number of outliers. The poor prediction performance when $V<-0.11$ is a consequence of a small portion ($0.78$ percent for all replications) of data in this range. 

\begin{figure*}[!h]
	\centering
	\includegraphics[width =0.85\linewidth]{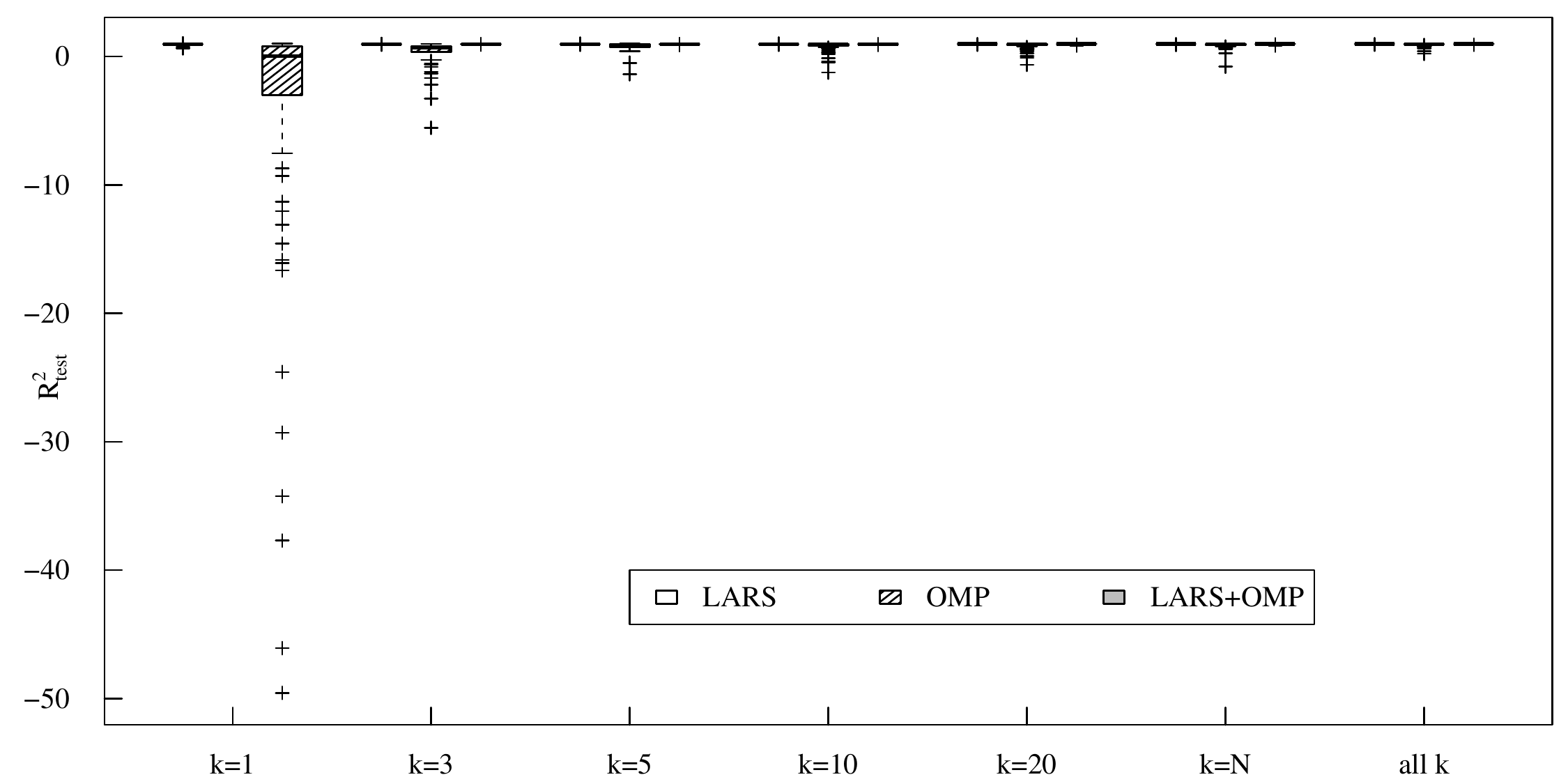}
	\caption{Truss deflection - box plots of $R_{\text{test}}^2$ using different values of $k$ with $50$ data points ($100$ replications).}
	\label{Truss_boxplot}
\end{figure*}

\begin{table}[!ht]
	\centering
	\begin{tabular}{l|*{3}{c}}
		\toprule
		&{LARS}&{OMP}&{LARS+OMP}\\
		\midrule
		$k=1$ & 0.9631 & -6.2248 &\\
		$k=3$ & 0.9651 & 0.3873 & 0.9641 \\
		$k=5$ & 0.9658 & 0.7915 & 0.9660 \\
		$k=10$ & 0.9692 & 0.8273 & 0.9693 \\
		$k=20$ & 0.9726 & 0.8721 & 0.9735 \\
		$k=N$ & 0.9735 & 0.8974 & 0.9741 \\
		all $k$ & 0.9744 & 0.9315 & 0.9762\\
		\bottomrule
	\end{tabular}
	\caption{Truss deflection - mean of $R_{\text{val}}^2$ with $50$ data points ($100$ replications).}
	\label{Truss_table}
\end{table}

\begin{figure}[!ht]
	\centering
	\includegraphics{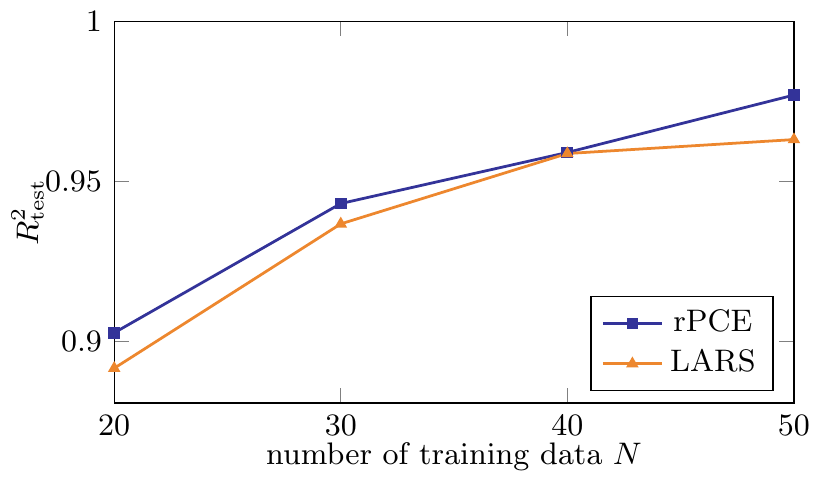}
	\caption{Truss deflection - mean of $R_{\text{val}}^2$ versus different values of $N$ ($100$ replications).}
	\label{TrussR2_N}
\end{figure}
Based on the validation data, $R^2_{\text{test}}$ is computed at each replication and the distribution of $R^2_{\text{test}}$ over $100$ replications is given in Fig.~\ref{Truss_boxplot}. The results with $k=1$ indicate the running of LARS and OMP with the whole set of data, thus no refinement of the basis by rPCE and ``all $k$" means that rPCE is run based on the combination of candidate polynomials generated with $k=[3,5,10,20,N]$. Although the performance of OMP is much enhanced with the application of rPCE, LARS is still better than OMP, whatever the value of $k$. The rPCE model combining LARS and OMP seems to have the same performance with the rPCE model based on LARS itself. Table~\ref{Truss_table} presents the associated mean of $R_{\text{test}}^2$. As seen, the highest mean appears with LARS+OMP when all $k$ values are considered, but, with the same configurations, the difference between LARS and LARS+OMP is only $0.0018$. Optimizing the selection of candidate polynomials at each replication, as displayed in Fig.~\ref{TrussR2_N}, the mean value reaches
$0.9770$ for the ``all $k$" option. The slight superiority of rPCE to LARS is also seen with $N=20,30,40$.

\begin{table}[!ht]
	\centering
	\caption{Truss deflection - mean of the total Sobol' indices with $50$ data points ($100$ replications).}
	\begin{tabular}{l|*{4}{c}}
		\toprule
		&Reference&{rPCE}&{LARS}&{OMP}\\
		\midrule
		$E_h$ & 0.367 & 0.3713 & 0.3748 & 0.4295\\
		$E_o$ & 0.010 & 0.0121 & 0.0135 & 0.2290\\
		$A_h$ & 0.388 & 0.3695 & 0.3715 & 0.4037\\
		$A_o$ & 0.014 & 0.0127 & 0.0135 & 0.2291\\
		$P_1$ & 0.004 & 0.0046 & 0.0057 & 0.2105\\
		$P_2$ & 0.031 & 0.0359 & 0.0365 & 0.2251\\
		$P_3$ & 0.075 & 0.0750 & 0.0759 & 0.2808\\
		$P_4$ & 0.079 & 0.0756 & 0.0751 & 0.2557\\
		$P_5$ & 0.035 & 0.0355 & 0.0361 & 0.2271\\
		$P_6$ & 0.005 & 0.0048 & 0.0061 & 0.1891\\
		\midrule
		$\sum$ & 1.008 & 0.9969 & 1.0086 & 2.6795\\
		\bottomrule
	\end{tabular}
	\label{table_SU_Truss_N50}
\end{table}

Global sensitivity analysis is conducted by computing the total Sobol' indices based on the PCE coefficients. The reference values listed in Table~\ref{table_SU_Truss_N50} are obtained with $5.5\times 10^6$ Monte Carlo simulations \cite{blatman2011adaptive}. Since the characteristics of the horizontal bars impact more the displacement at midspan than the oblique ones, the total Sobol' indices of $E_h$ and $A_h$ are much larger than those of $E_o$ and $A_o$. Moreover, due to the same type of probabilistic distribution and the fact that the products $E_hA_h$ (resp. $E_oA_o$) are the physically meaningful quantities in the analysis, $E_h$ and $A_h$ (resp. $E_o$ and $A_o$) have similar importance to the response. Considering the variables of $P_i$, $i=1,\ldots,6$, $P_i$ and $P_{7-i}$ play the same role due to the geometric symmetry of the structure and greater sensitivities are observed for loads closer to the midspan. The above conclusions are clearly supported by the estimations of rPCE and LARS. In contrast, the largely biased estimation by OMP might give a wrong understanding of the physical phenomena. For instance, one may falsely conclude that the actually negligible interactions among inputs have great effects on the midspan deflection, since the sum of the total Sobol' indices obtained by OMP is much larger than $1$. 

\begin{figure*}[!ht]
	\centering
	\includegraphics[width =0.85\linewidth]{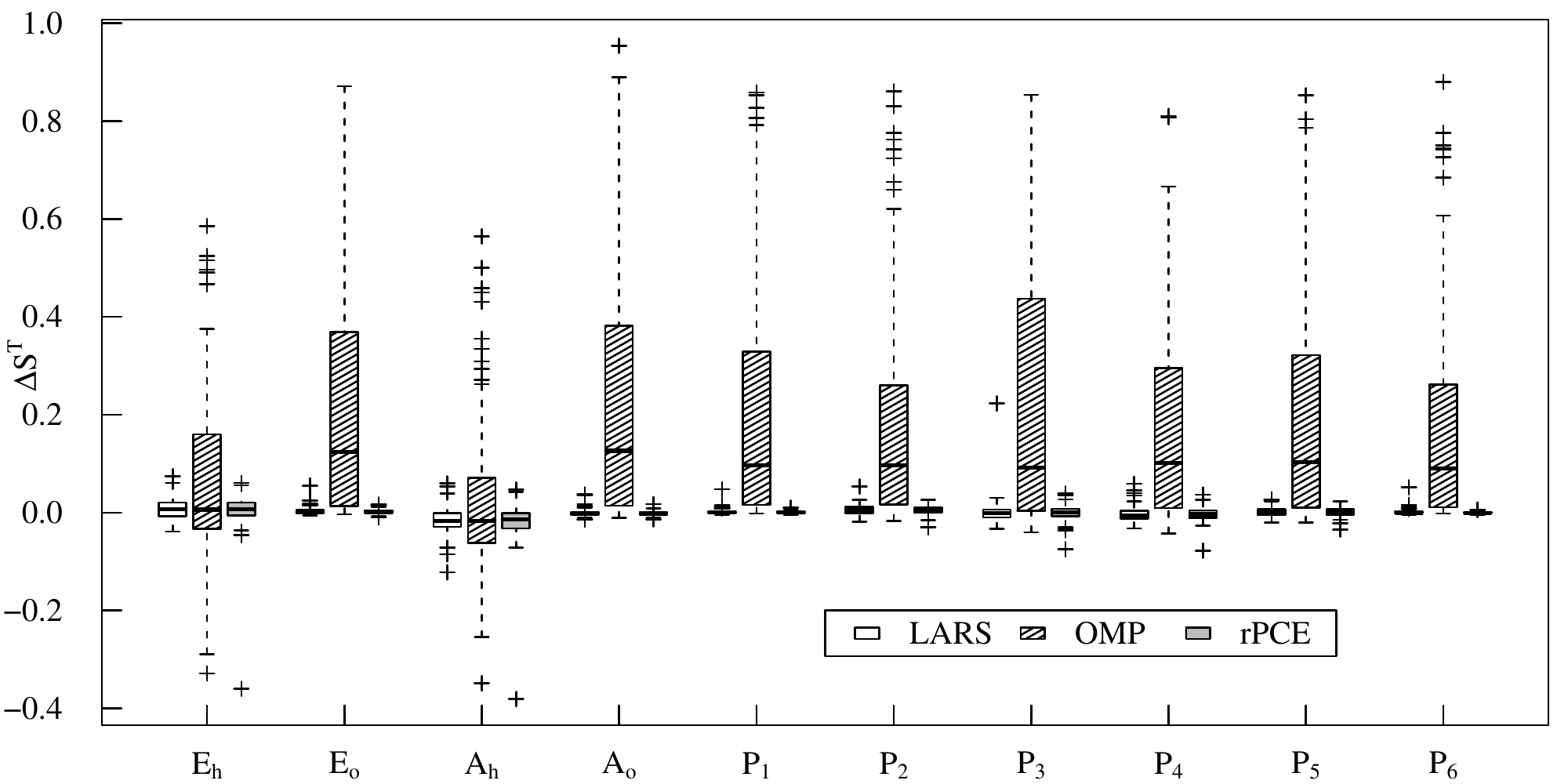}
	\caption{Truss deflection - the estimation error of total Sobol' indices with $50$ data points ($100$ replications).}
	\label{SU_Truss_N50}
\end{figure*}

The distribution of the prediction error of total Sobol' indices $\Delta S^T$ is given in Fig.~\ref{SU_Truss_N50}. In addition to the largely biased and scattered OMP, rPCE and LARS has similar $\Delta S^T$ distribution with relatively small variances.

\subsection{Estimation of specific absorption rate}
\label{subsec:SAR}
\begin{figure}[!ht]
	\centering
	\includegraphics[width=0.5\linewidth]{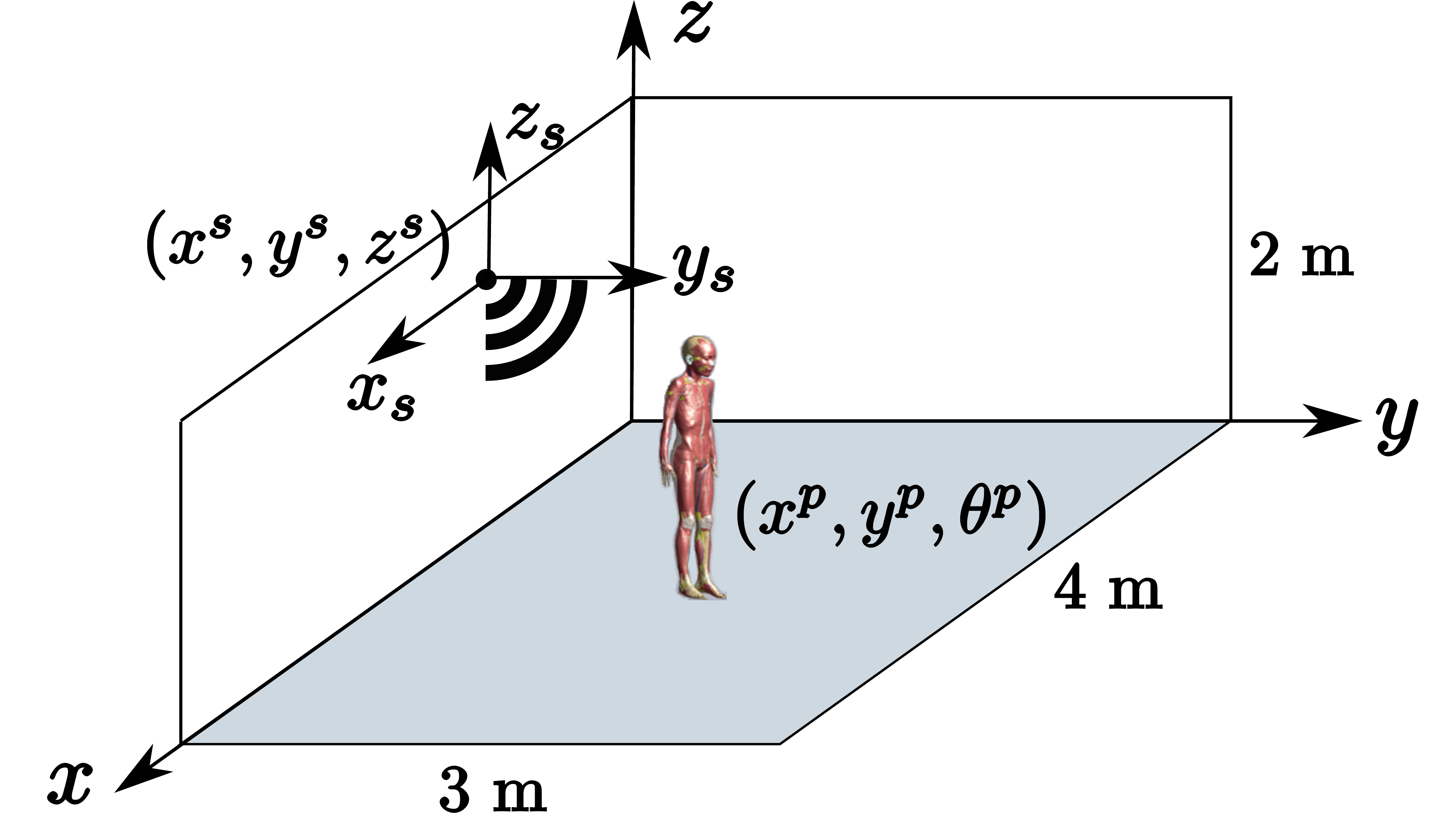}
	\caption{Sketch of the human-exposure estimation in an indoor down-link scenario.}
	\label{sketchExposure}
\end{figure}

The population is surrounded by a increasing number of wireless local area networks (WLAN) and the electromagnetic exposure of human body by WLAN access points needs to be estimated to make sure the exposure level is under the limit \cite{van2011research}. Here, an indoor down-link scenario is considered, as sketched in Fig.~\ref{sketchExposure}. A high-resolution model of a 8-year girl ($1.36$ m high), named as ``Eartha", from the Virtual Classroom \cite{gosselin2014development}, is standing inside a $4 \times 3 \times 2$ m$^3$ room, which is equipped with a WLAN source operating at $2.4$ GHz. The field emitted by the source is measured using the StarLab near-field-measurement system, which is based on spherical wave expansion \cite{hald1988spherical}, by Microwave Vision Group (MVG$^{\textregistered}$). With an in-house finite-difference-time-domain (FDTD) code, the whole-body  specific absorption rate (SAR) \cite{liorni2016exposure}, which is the system response here, is computed as the ratio of the total power absorbed in the body to the mass of the human model and with the unit mW/kg. 

The parameters considered are the position of the emitting source and the human model, whose coordinates are denoted by $(x^s,y^s,z^s)$ and $(x^p,y^p,z^p)$, respectively. $z^p$ is set as $0$, since we consider that the human model is standing on the ground. The human orientation $\theta^p$, which is defined as the angle between the direction faced by the human model and $x$-axis, may matter and is taken into account.The reflection by the walls, ceiling and ground is neglected in the simulation and the WLAN source is attached to the walls. Thus, six parameters are involved. $x^s$, $y^s$, $z^s$, $x^p$, $y^p$ are assumed to be uniformly distributed over $[0.3, 3.7]$, $[0.3, 2.7]$, $[0.25,2]$, $[0.05,3.95]$, $[0.05,2.95]$ in meters and $\theta^p$ over $[0,360)$ in degrees, where the lower bound value $0.3$ m is the minimum distance between the human model and the wall, $0.25$ m is the minimum height of the source and $0.05$ m is the minimum distance of the WLAN source to the wall. 

The number of input variables can be reduced via a coordinate transformation. Without the reflection by the walls, the system response is actually driven by the relative position between the source and the human model. The relative position is represented in the $(x,y)$ plane. In the local coordinate system of the source, as shown in Fig.~\ref{sketchExposure}, position and orientation of the human model are denoted by polar coordinates $(r_s^p,\phi_s^p)$ and $\theta_s^p$. Thus, four parameters $r_s^p$, $\phi_s^p$, $\theta_s^p$, and $z^s$ are used in the following uncertainty analysis.

\begin{figure}[!ht]
	\centering
	\subfigure[rPCE ($R_{\text{test}}^2=0.9102$)]{\includegraphics[width = 0.33\linewidth]{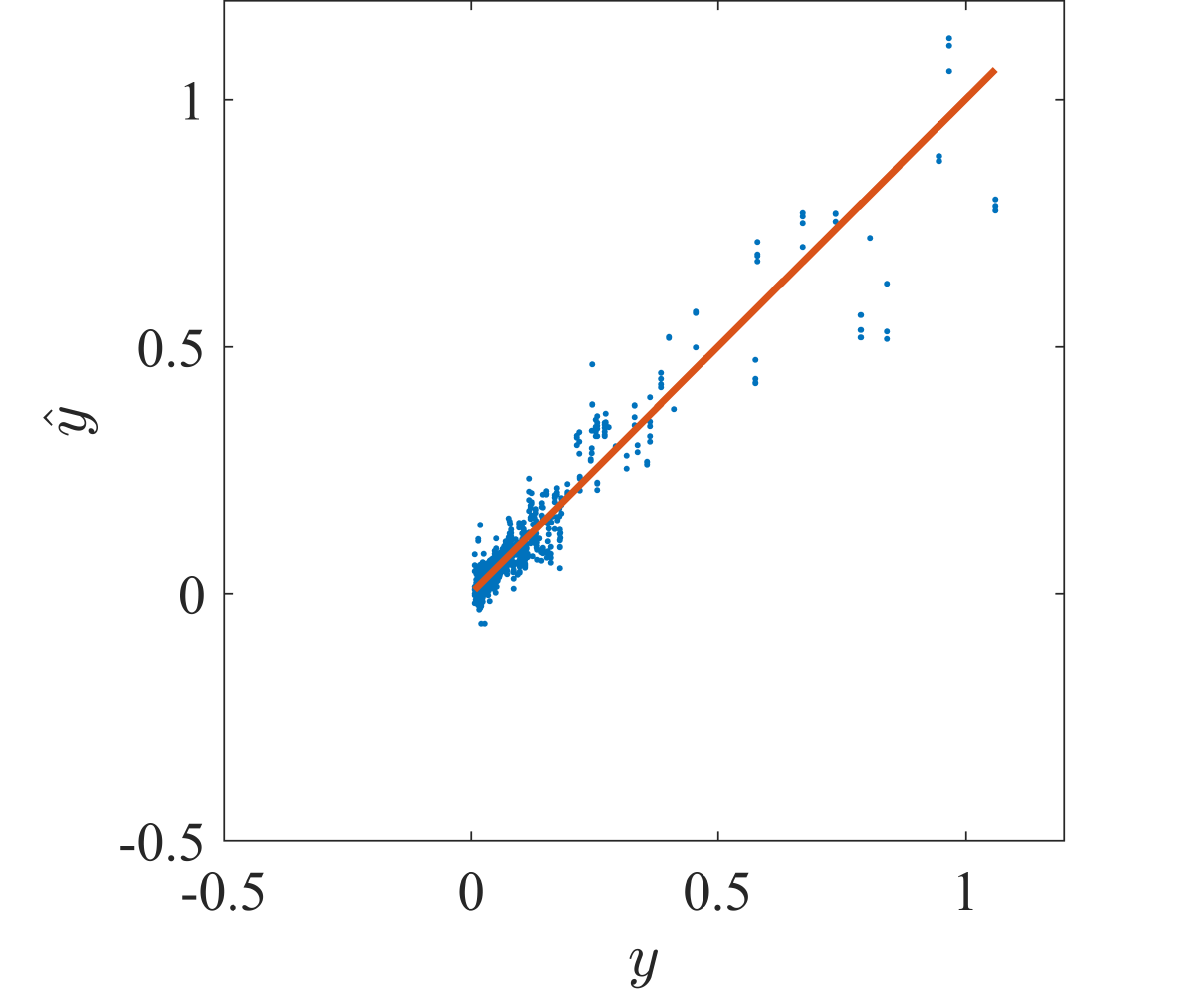}\label{predSAR_rPCE}}~
	\subfigure[LARS ($R_{\text{test}}^2=0.8688$)]{\includegraphics[width = 0.33\linewidth]{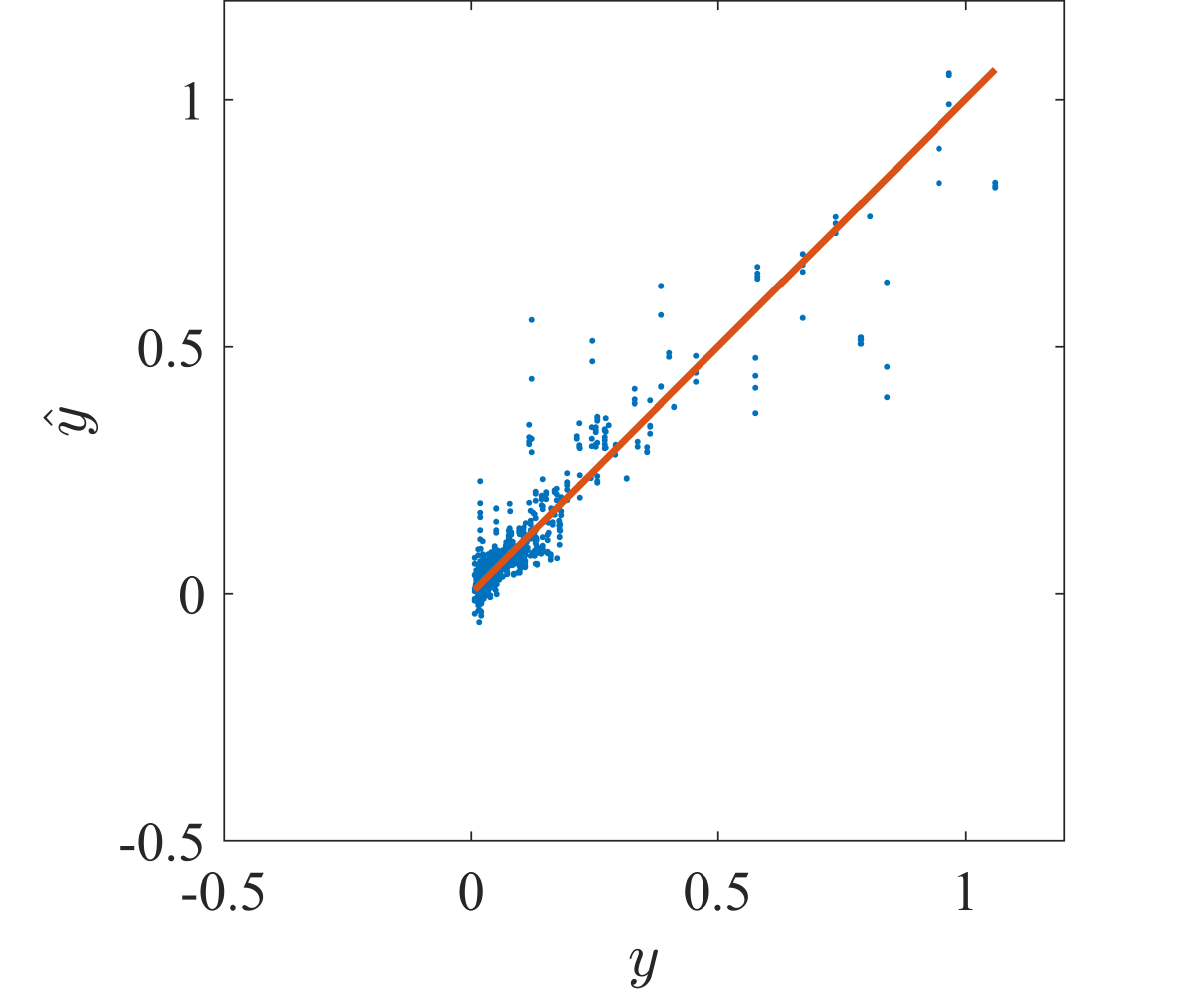}\label{predSAR_LARS}}~
	\subfigure[OMP ($R_{\text{test}}^2=0.7269$)]{\includegraphics[width = 0.33\linewidth]{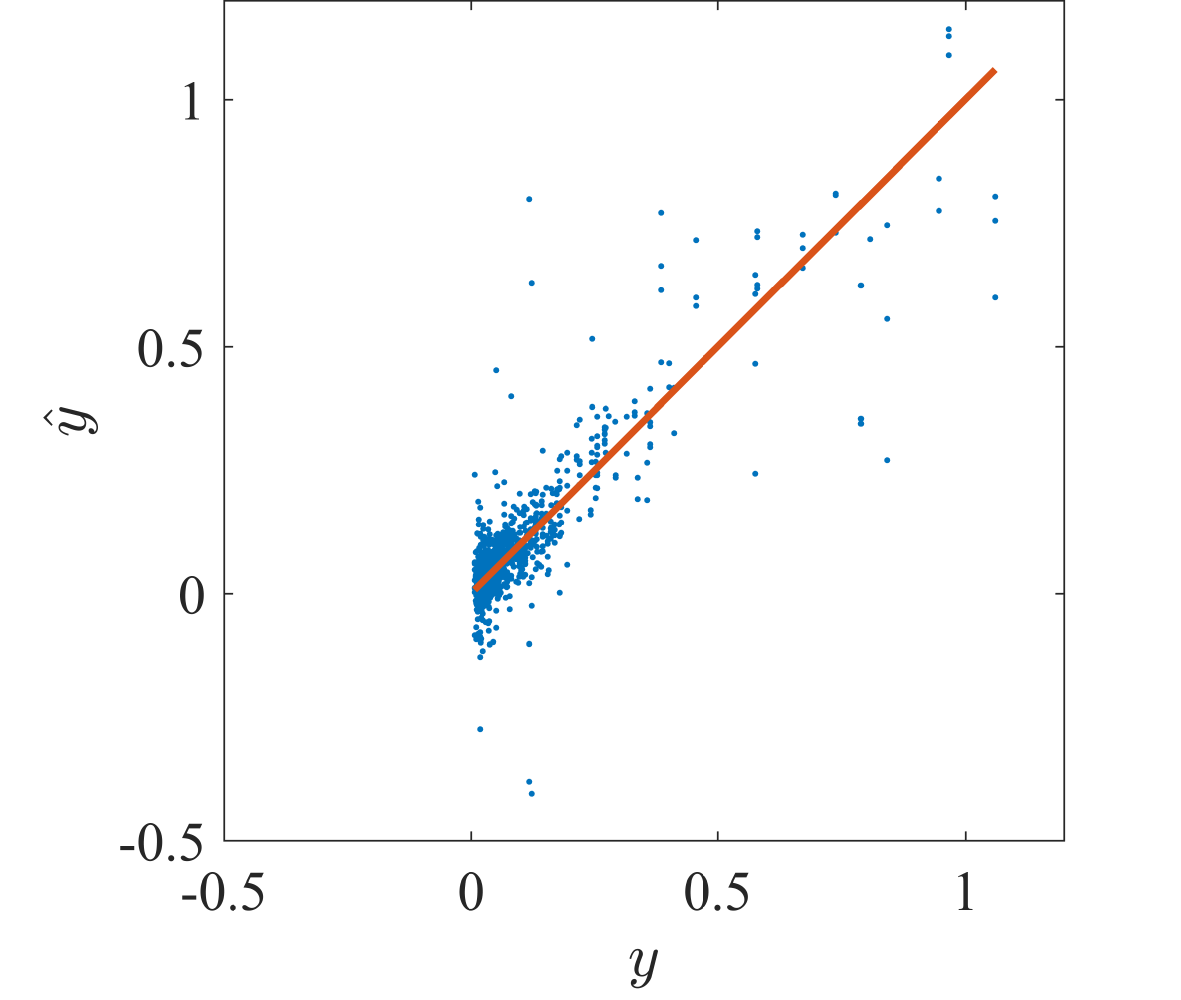}\label{predSAR_OMP}}
	\caption{SAR estimation - prediction of validation data by (a) rPCE, (b) LARS and (c) OMP with $340$ data points ($100$ replications).}
	\label{predSAR}
\end{figure}

\begin{figure*}[!ht]
	\centering
	\includegraphics[width =0.85\linewidth]{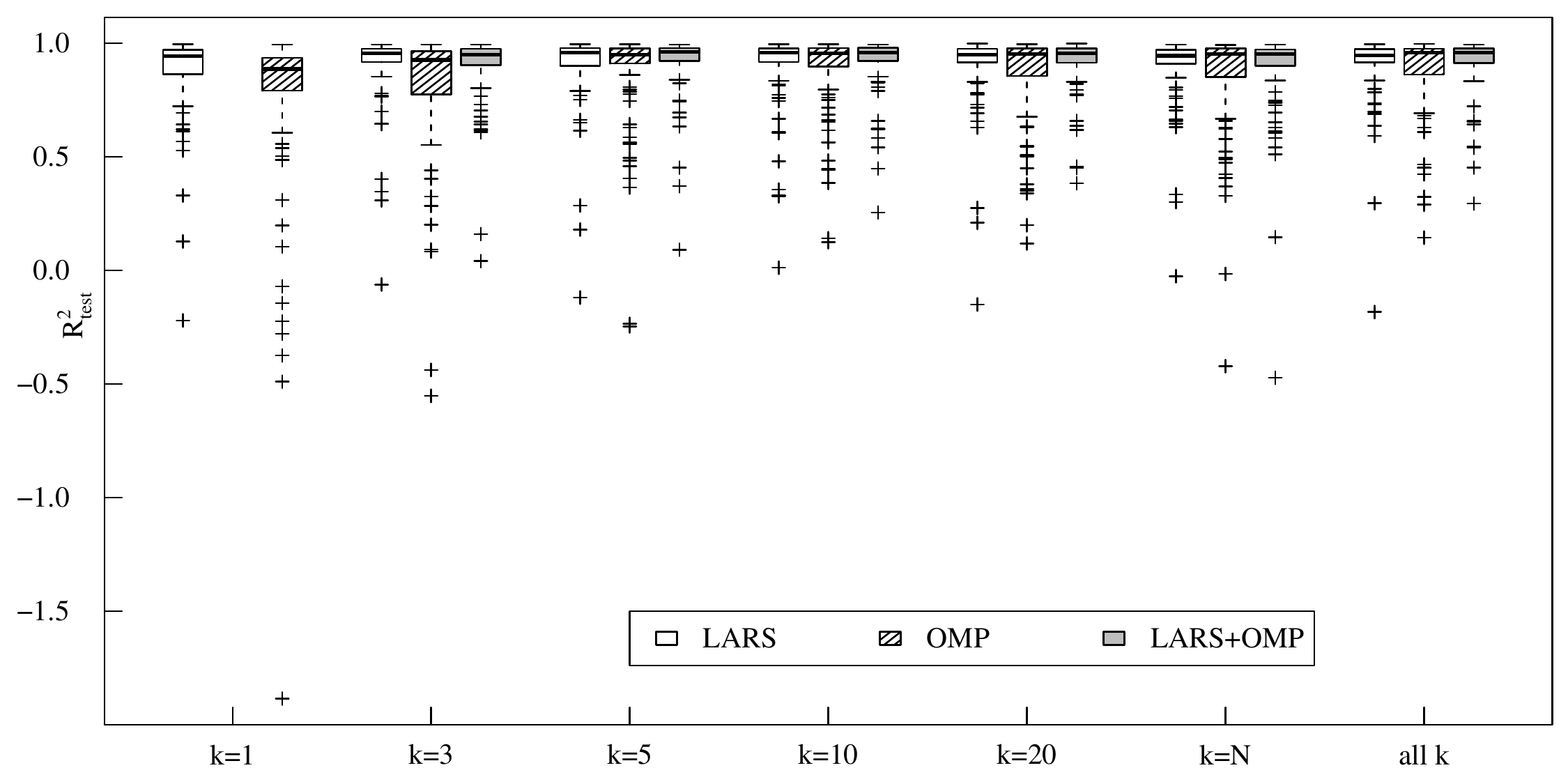}
	\caption{SAR estimation - box plots of $R_{\text{test}}^2$ with different values of $k$ ($100$ replications).}
	\label{SAR_boxplot}
\end{figure*}

Sampling $350$ points from the input space with the Latin-Hypercube sampling method, the prediction performance of the obtained surrogate models is estimated with the leave-many-out approach, where $10$ data are randomly chosen from the experimental design for validation and an approximation of $R_{\text{test}}^2$ is yielded by repeating this process $100$ times. Consequently, with the remaining $340$ data, surrogate models are obtained with LARS, OMP and rPCE. Then, a validation set of size $10^3$ is computed and the results are shown in Fig.~\ref{predSAR}. As seen, the whole-body SAR is smaller than $0.2$ for most of cases ($90$ percents for all replications) in this scenario. However, the three approaches can provide unbiased estimations when the SAR value is larger than $0.2$, in addition to the the superiority of rPCE to LARS and OMP in variance and suppression of outliers. 
The associated box plots of $R_{\text{test}}^2$ is given in Fig.~\ref{SAR_boxplot}. The refinement by rPCE reduces the variance of modeling by LARS and OMP with different values of $k$, except for the case with OMP and $k=3$. The combination of LARS and OMP seems to be the best option for rPCE and actually is selected by the suggested scheme in Section \ref{subsec:rPCE_and} during all replications (although three options are available at each replication), since LARS has the same-level performances with OMP. Table~\ref{SAR_table} shows the mean of $R_{\text{test}}^2$. 
\begin{table}[!ht]
	\centering
	\begin{tabular}{l|*{3}{c}}
		\toprule
		&{LARS}&{OMP}&{LARS+OMP}\\
		\midrule
		$k=1$ & 0.8799 & 0.7500 &\\
		$k=3$ & 0.9085 & 0.8186 & 0.9046 \\
		$k=5$ & 0.9067 & 0.8771 & 0.9182 \\
		$k=10$ & 0.8995 & 0.8854 & 0.9171 \\
		$k=20$ & 0.9033 & 0.8628 & 0.9157 \\
		$k=N$ & 0.8995 & 0.8521 & 0.8893 \\
		all $k$ & 0.9068 & 0.8794 & 0.9178\\
		\bottomrule
	\end{tabular}
	\caption{SAR estimation - mean of $R_{\text{val}}^2$ with $340$ data points ($100$ replications).}
	\label{SAR_table}
\end{table}

\begin{table}[!ht]
	\centering
	\begin{tabular}{l|*{3}{c}}
		\toprule
		&{rPCE}&{LARS}&{OMP}\\
		\midrule
		$r_s^p$ & 0.9809 & 0.9714 & 0.9761\\
		$\phi_s^p$ & 0.0128 & 0.0357 & 0.0984\\
		$z^s$ & 0.2175 & 0.1954 & 0.2925\\
		$\theta_s^p$ & 0.0098 & 0.0316 & 0.0743\\
		\midrule
		$\sum$ & 1.2210 & 1.2341 & 1.4412 \\
		\bottomrule
	\end{tabular}
	\caption{SAR estimation - mean of the total Sobol' indices with $340$ data points ($100$ replications).}	
	\label{table_SU_SAR_N340}
\end{table}

\begin{figure}[!ht]
	\centering
	\subfigure[]{\includegraphics[width = 0.45\linewidth]{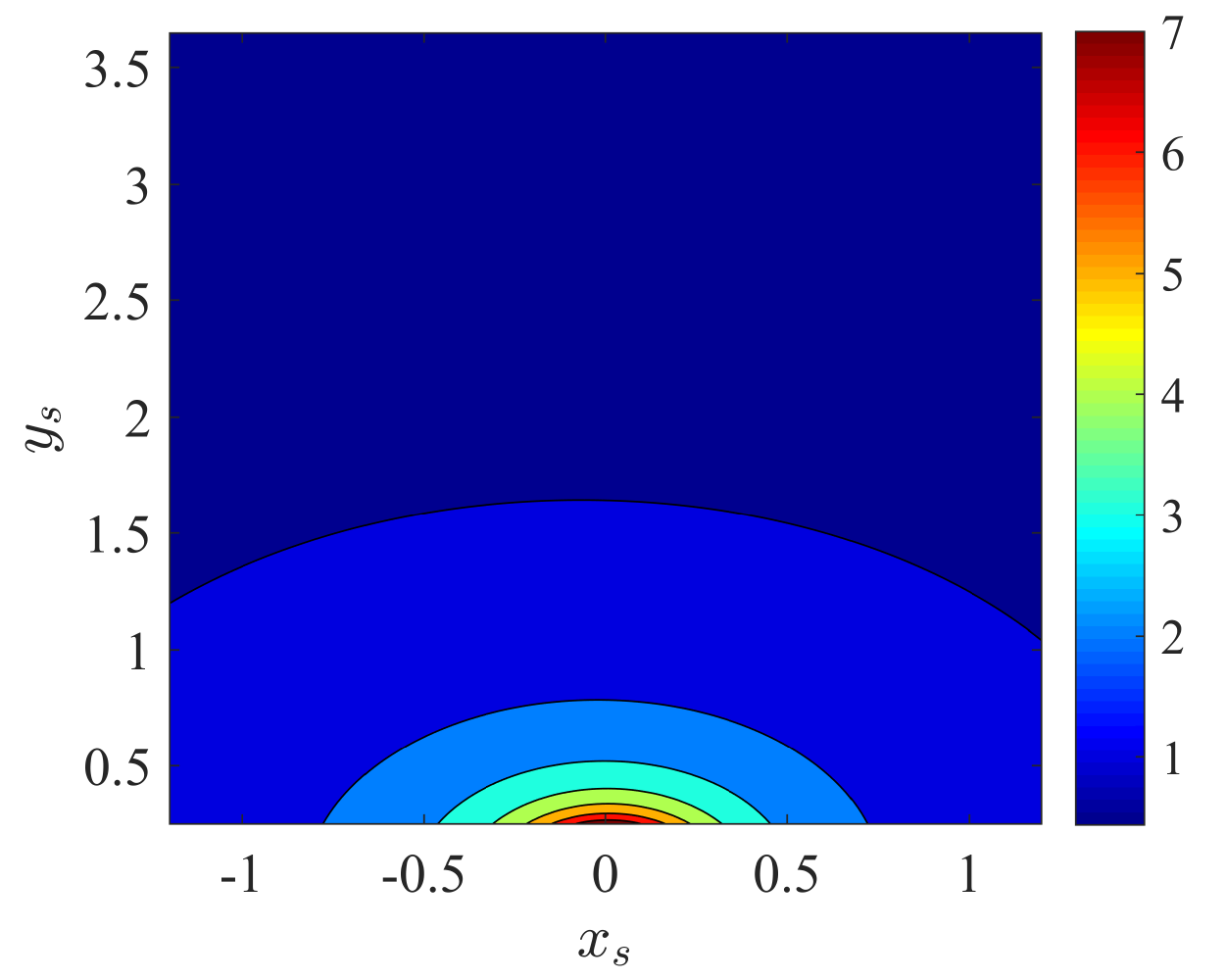}\label{fieldMapCart}}~
	\subfigure[]{\includegraphics[width = 0.45\linewidth]{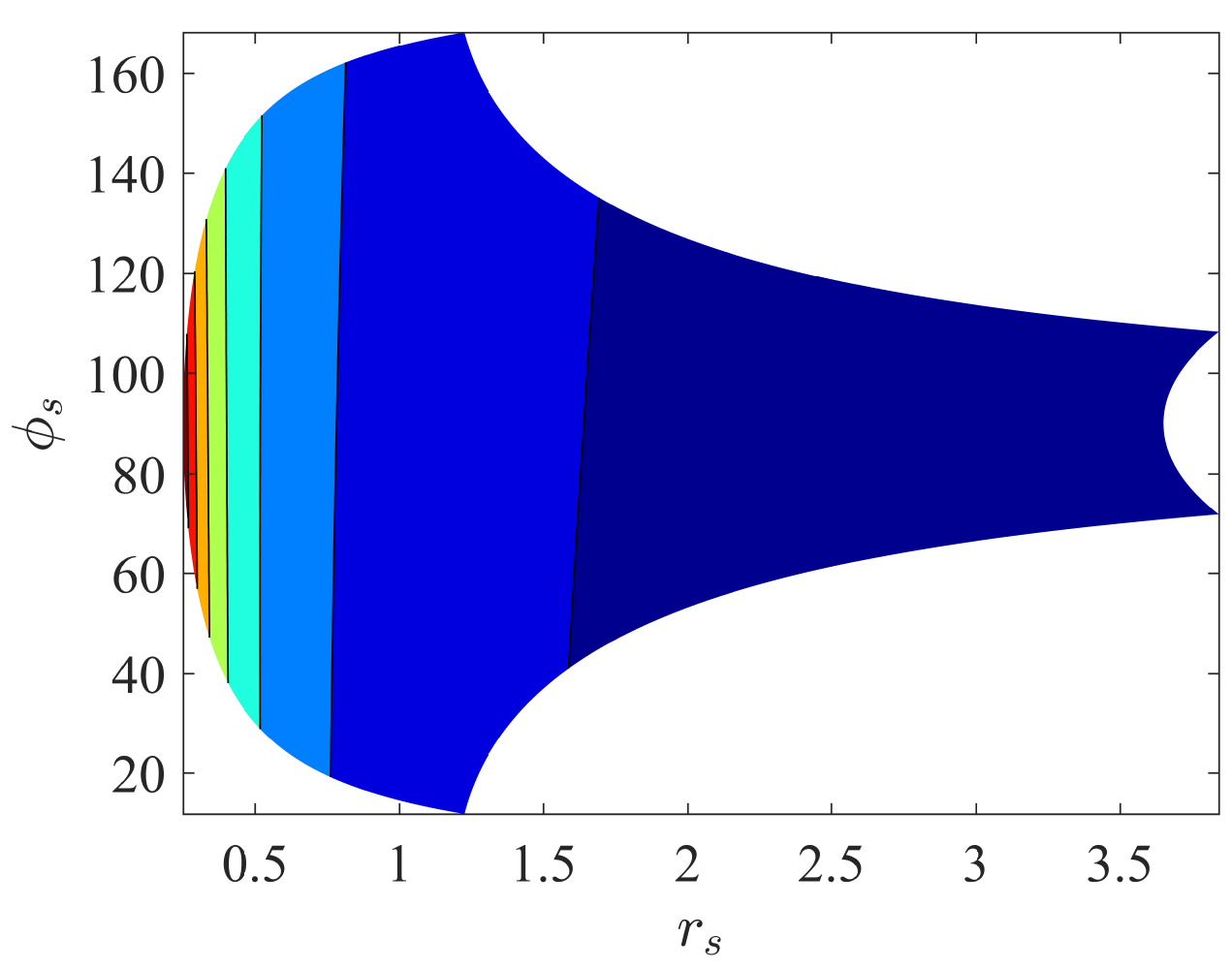}\label{fieldMapPolar}}
	\caption{SAR estimation - contour of electric-field intensity (a) in the $(x,y)$ plane and (b) its representation in the polar coordinate system, $z_s=0$.}
	\label{fieldMap}
\end{figure}
\begin{figure}[!ht]
	\centering
	\includegraphics[width =0.7\linewidth]{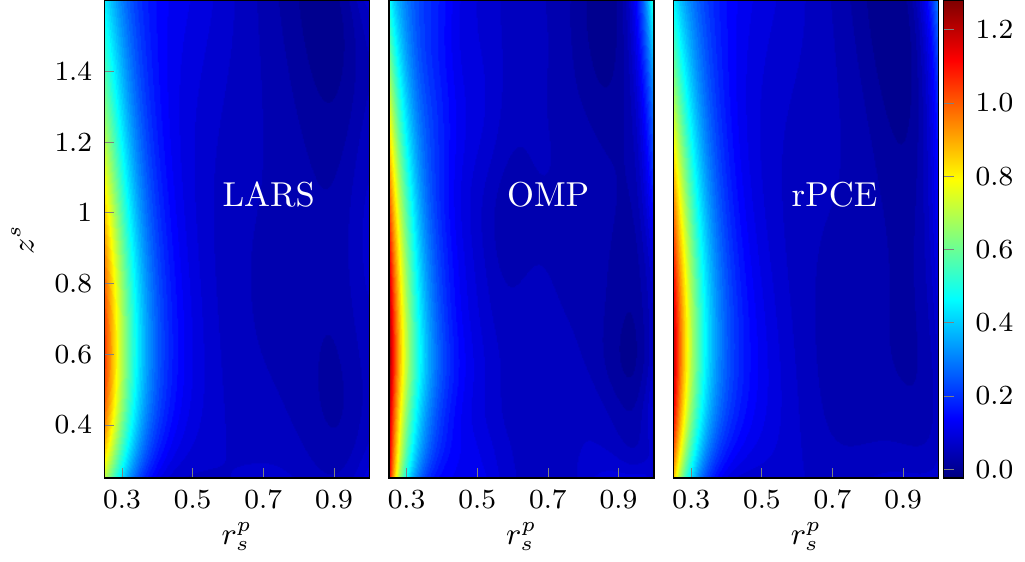}
	\caption{SAR estimation - the prediction of whole-body SAR with $340$ data points ($100$ replications).}
	\label{predictSAR}
\end{figure}

The total Sobol' indices are computed based on the PCE coefficients and the mean values are presented in Table~\ref{table_SU_SAR_N340}. As seen, the whole-body human exposure is mainly impacted by the relative distance $r_s^p$ and the height of the source $z^s$ has a smaller influence. The small value w.r.t. the relative angle between the human model and the source, $\phi_s^p$, might be explained by looking at the contours of electric-field intensity in Fig.~\ref{fieldMap}, where the WLAN source locates at the center of a wall and field values are sampled in the $(x_s,y_s)$ plane with $z_s=0$. As observed, the dependency of wave strength on radiation directions is weak. The human orientation $\theta_s^p$ affects the distribution of SAR in the human body. However, as the mean value of this distribution, the whole-body SAR is not much affected by $\theta_s^p$. The sum of the total Sobol' indices in Table~\ref{table_SU_SAR_N340} is larger than $1$ and the excess values indicate that $z^s$ impacts the response mainly through its interaction with $r_s^p$.  Such an interaction can be viewed from the map of predicted SAR in Fig.~\ref{predictSAR}, where $\phi_s^p$, $\theta_s^p$ are fixed to zero and $r_s^p$, $z^s$ are uniformly sampled over $[0.25,1]$, $[0.25,2]$, respectively. The amplitude of each pixel in the map is a mean of $100$ predictions by the built PCE models during all replications. The three approaches provide similar results. 

\begin{figure*}[!ht]
	\centering
	\includegraphics[width =0.85\linewidth]{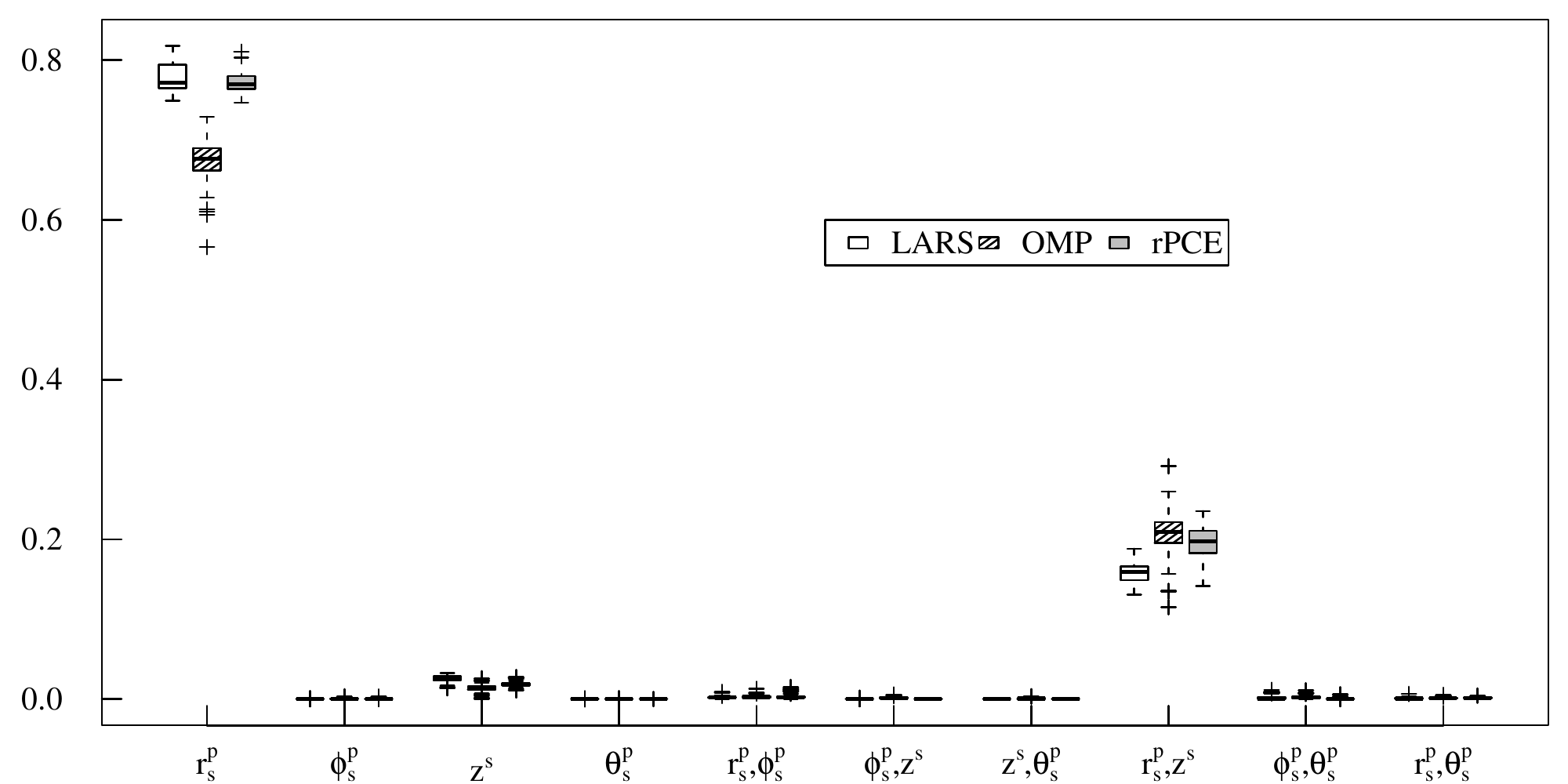}
	\caption{SAR estimation - the estimation of first-order and second-order Sobol' indices with $340$ data points ($100$ replications).}
	\label{SU_SAR_N340}
\end{figure*}
Considering the height of the human model is $1.36$ m, tissues mainly locate at the heights between $0.4$~m $\le z^s \le 1.0$~m. One observes that the whole-body SAR is rather small when the source is farther from this influential region of the human model. $r_s^p$ and $z^s$ model the distance between the source and this influential region together and their interactions happen. The distribution of the estimated first-order and second-order Sobol' indices is proposed in Fig.~\ref{SU_SAR_N340}, which presents that $r_s^p$ and its interaction with $z^s$ contribute the most to the uncertainty of the response.

\subsection{Example with varied input dimension}
{To investigate the effects of the dimension of input on the modeling performance, the following test function \cite{marelli2014uqlab} is used,
	\begin{equation}
	y = 3+\frac{1}{M}\sum_{k=1}^{M}k(x_k^3-5x_k)+\ln\left(\frac{1}{3M}\sum_{k=1}^{M}k(x_k^2+x_k^4)\right)+x_1x_2^2-x_3x_5+x_2x_4+x_{M-4}+x_{M-4}x_M^2, 
	\end{equation}
	where $M$ denotes the number of variables, which are independent and uniformly distributed in the range of $[1,2]$. To increase the non-linearity, the range of $x_{20}$ (when $M\ge 20$) is changed as $[1,3]$.
	
	The value of $M$ changes from $11$ to $41$ with a step $5$ and the size of experimental design $N$ is fixed as $200$ independently of $M$. For a statistical assessment of the modeling performance (accuracy and efficiency), $50$ replications are performed and $10^3$ data are used for the independent test at each replication. Remark that, due to randomness of the LHS method, different training and test datasets are used in replications.
	
	From the methodology of rPCE, one knows that the computational cost is proportional to the number of resampled datasets. In Section ~\ref{subsec:resampling}, the suggested (not obliged) setting of $k$ is a set of values, i.e., $k=\{3,5,10,20,N\}$. As a result, the corresponding computational cost would be high when the ED size $N$ is large. Here, a lighter setting of $k$, $k=\{3,5,10,20\}$, is applied and improved modeling performances are still observed as presented by the following results. For the configuration of UQLab, which the running of LARS and OMP is based on, the maximum value of total degree $p$ is set as $5$.}

\begin{figure}[!ht]
	\centering
	\includegraphics[width=0.85\linewidth]{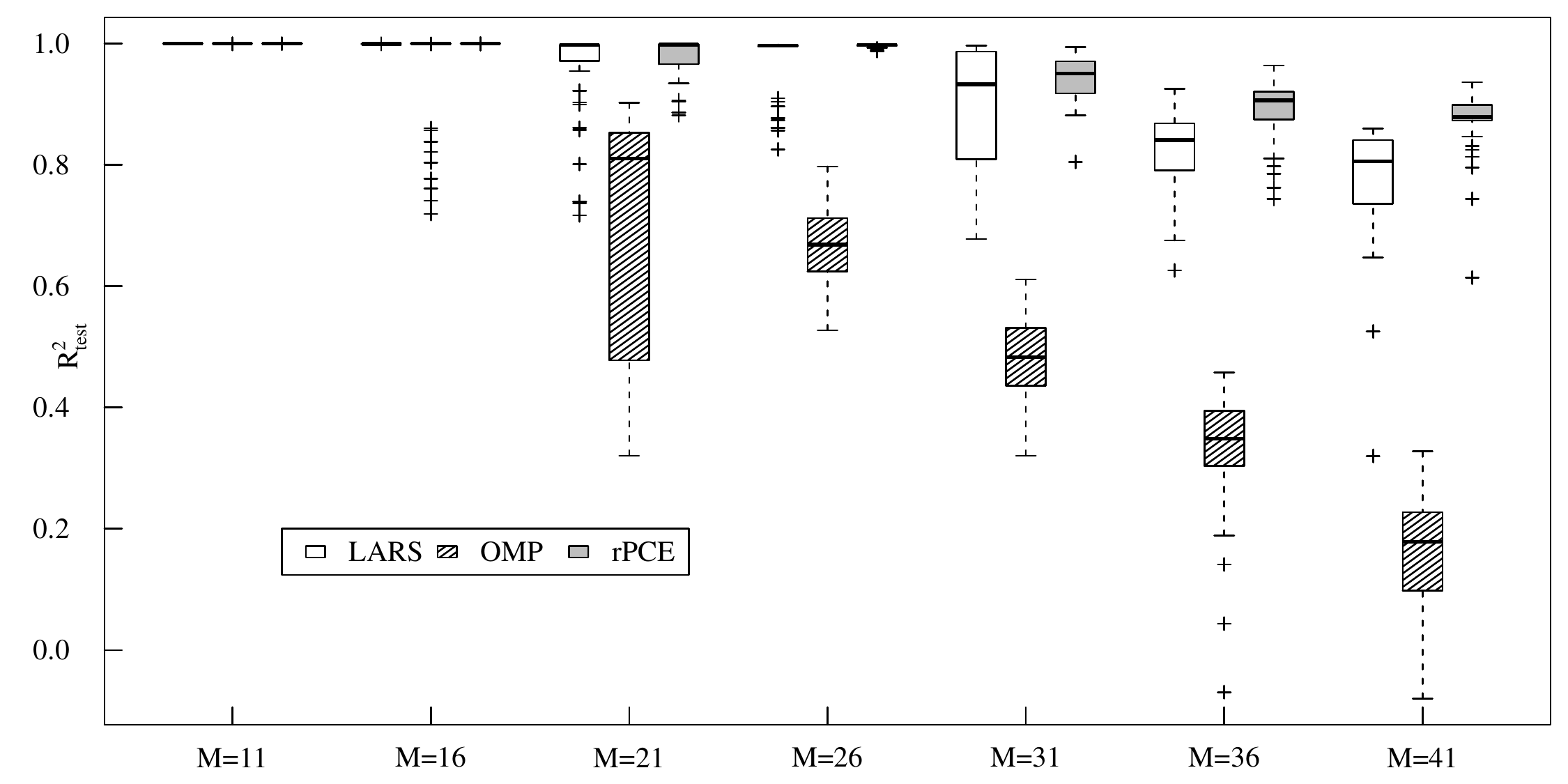}
	\caption{Example with varied dimension - box plots of $R_{\text{test}}^2$ with different values of $M$ ($50$ replications).}
	\label{boxplot_R2_highDimen}
\end{figure}
{The effects of $M$ on the modeling accuracy can be observed from the distribution of $R^2_{\text{test}}$ in Fig.~\ref{boxplot_R2_highDimen}. When $M$ equals $11$ and $16$, accurate models are constructed with the three approaches, although outliers appear with OMP and $M=16$. As the dimension of input increases, the modeling accuracy becomes poorer in both variance and bias. While LARS performs much better than OMP when $M\ge 21$, substantial advantages of rPCE (with suggested configurations) are observed. When $M\in\{31,36, 41\}$, the mean value of $R^2_{\text{test}}$ with rPCE is larger than the value with LARS and OMP, and the variance of rPCE is also superior to the other two approaches.}

\begin{table}[!ht]
	\caption{Example with varied dimension - mean of $R_{\text{test}}^2$ with varied values of $k$ and $M$ ($50$ replications), ``L+O" denoting the combination of LARS and OMP.}\label{highDimen_table}
	\begin{tabular}{l|c|*{7}{c}}
		\toprule
		&&{M=11}&{M=16}&{M=21}&{M=26}&{M=31}&{M=36}&{M=41}\\
		\hline
		\multirow{2}{3em}{$k=1$}&LARS &0.9998&0.9995&0.9573&0.9679&	0.8985&	0.8260&0.7761\\
		&OMP&0.9998&0.9634&0.6940&0.6679&0.4832&0.3308&0.1536\\
		\hline
		\multirow{3}{3em}{$k=3$}&LARS &0.9997&0.9996&0.9422&0.9249&0.8646&0.8322&0.8125\\
		&OMP&0.9998&0.8072&0.7810&0.7737&0.6514&0.5358&0.3870\\
		&L+O&0.9998&0.9996&0.8929&0.8771&0.7810&0.7262&0.6805\\
		\hline
		\multirow{3}{3em}{$k=5$}&LARS &0.9998&0.9995&0.9600&0.9726&0.8899&0.8574&0.8351\\
		&OMP&0.9999&0.9552&0.8171&0.7915&0.6935&0.5894&0.4826\\
		&L+O&0.9999&0.9996&0.9511&0.9651&0.8630&0.8110&0.7681\\
		\hline
		\multirow{3}{3em}{$k=10$}&LARS &0.9999&0.9995&0.9714&0.9945&0.9316&0.8724&0.8445\\
		&OMP&0.9999&0.9963&0.8395&0.8194&0.7252&0.6239&0.5340\\
		&L+O&0.9999&0.9998&0.9668&0.9937&0.9210&0.8557&0.8193\\
		\hline
		\multirow{3}{3em}{$k=20$}&LARS &0.9999&0.9995&0.9824&0.9971&0.9523&0.8947&0.8714\\
		&OMP&0.9999&0.9999&0.8392&0.8195&0.7197&0.6149&0.5165\\
		&L+O&0.9999&0.9999&0.9784&0.9971&0.9404&0.8692&0.8391\\
		\hline
		\multirow{3}{3em}{all $k$}&LARS &0.9999&0.9996&0.9765&0.9965&0.9437&0.8904&0.8725\\
		&OMP&0.9999&0.9987&0.8423&0.8248&0.7316&0.6191&0.5011\\
		&L+O&0.9999&0.9998&0.9738&0.9961&0.9371&0.8790&0.8604\\
		\bottomrule
	\end{tabular}
\end{table}
{From the mean value of $R^2_{\text{test}}$ (w.r.t. $50$ replications) in Table~\ref{highDimen_table}, one finds the effects of varied $M$ on rPCE with different configurations. Remark that the cases with $k=1$ correspond with the modeling results based on LARS or OMP without the refinement by rPCE and the other cases are results of rPCE with different configurations. ``all $k$" here means the combination of selection results with $k=\{3,5,10,20\}$. As seen, when $M\ge 21$ and $k$ is configured as $10$, $20$, or ``all $k$", significant improvements are observed. When only OMP is applied, the mean value of $R^2_{\text{test}}$ increases from $0.6940$ to $0.8423$ when $M=21$. With LARS, the mean value increases from $0.7761$ to $0.8725$ when $M=41$. Remark that these two peak values are reached with the configuration of ``all $k$". Concerning on the source of candidate polynomials, since LARS performs much better than OMP, rPCE makes use of the selection results by LARS more often than OMP or their combination. Consequently, rPCE with the suggested setting (denoted by ``L+O" in Table~\ref{highDimen_table}), is slightly (not significantly) inferior to the setting only based on LARS in modeling accuracy.}  

\begin{figure}[!ht]
	\centering
	{\subfigure[$M=31$]{\includegraphics[width = 0.5\linewidth]{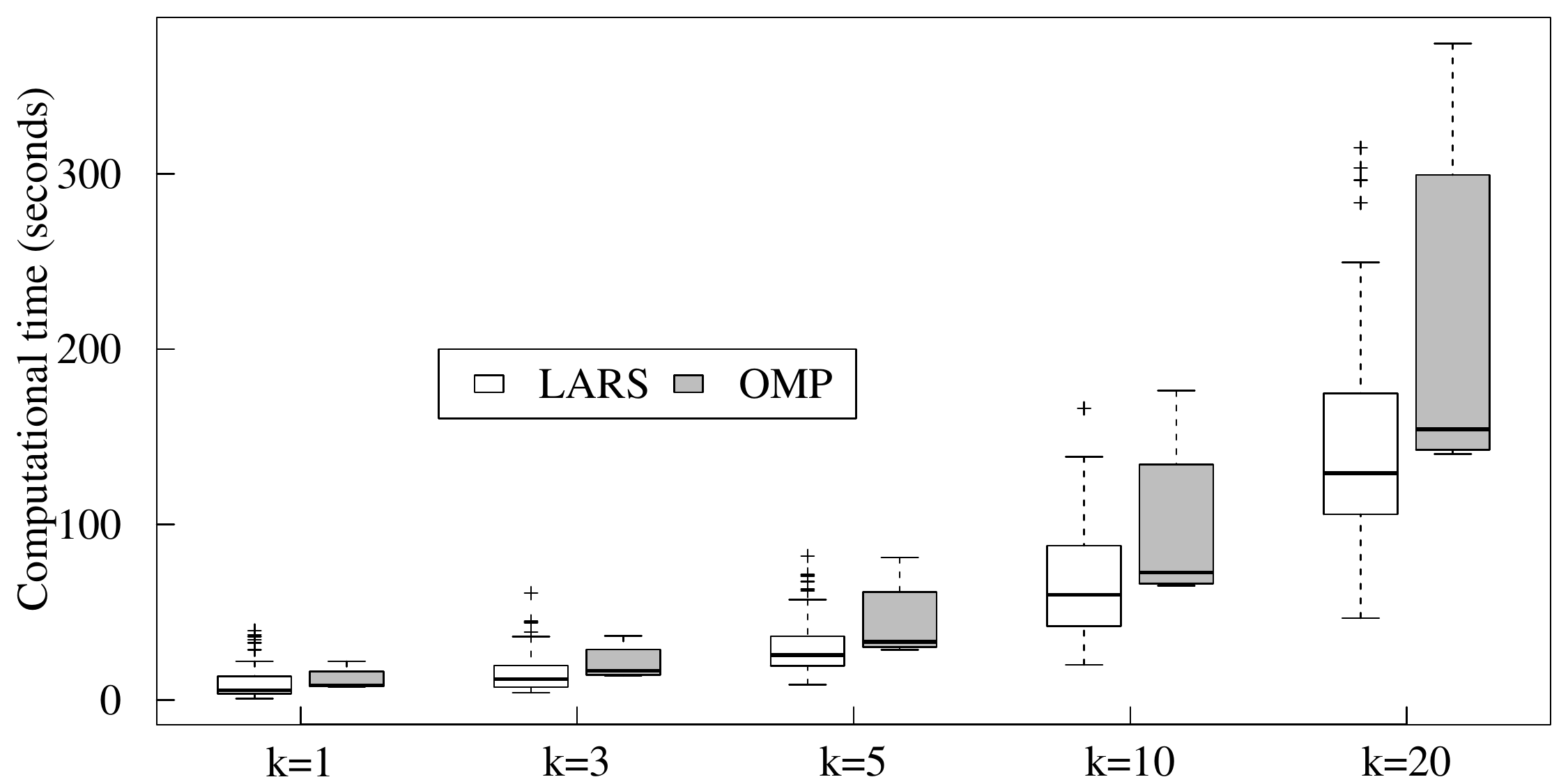}\label{timeBoxplot_M31}}~
		\subfigure[$k=10$]{\includegraphics[width = 0.5\linewidth]{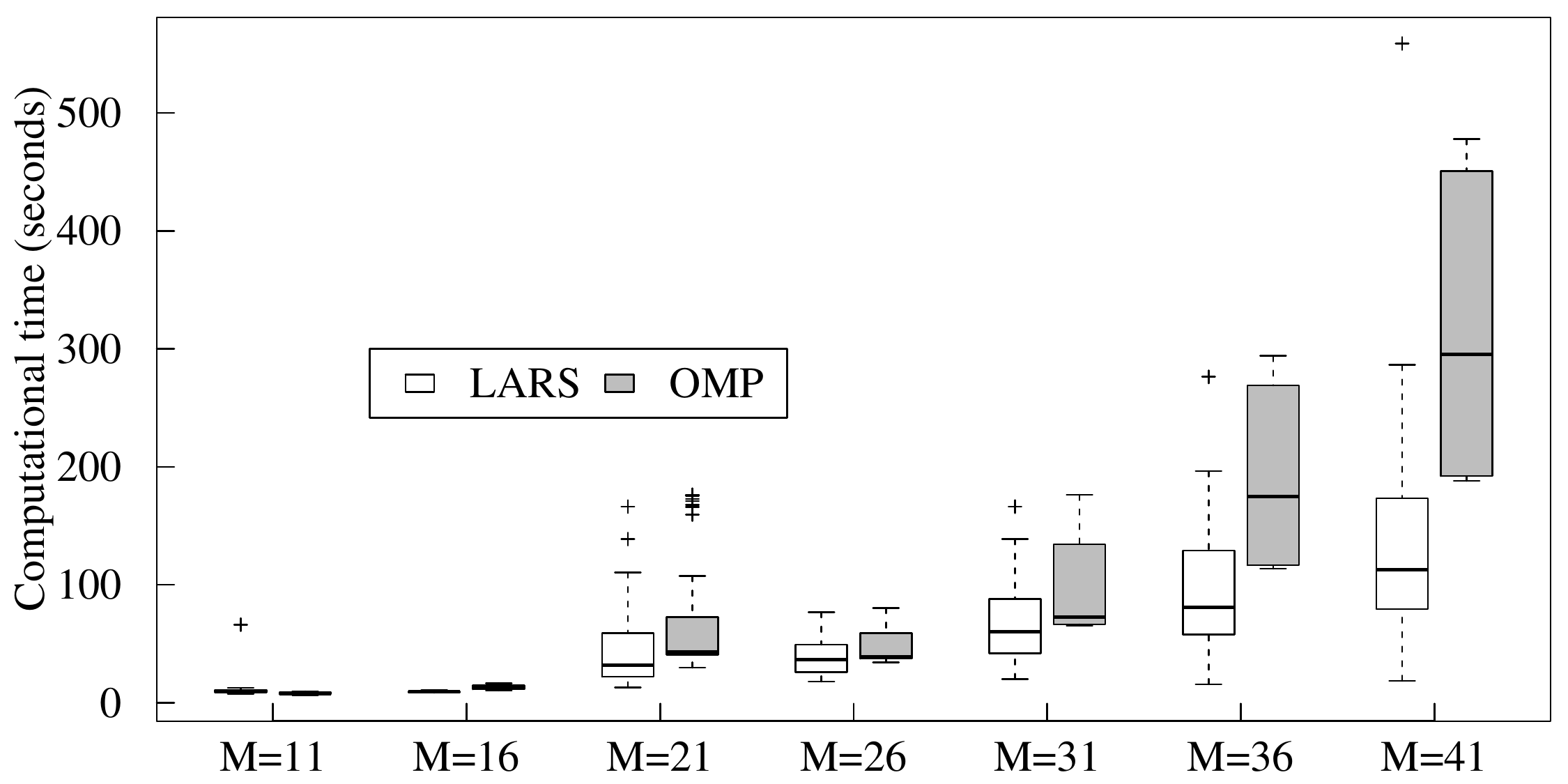}\label{timeBoxplot_k10}}}
	\caption{Example with varied dimension - computational time versus varied values of $k$ and $M$ ($50$ replications).}
	\label{timeBoxplot}
\end{figure}  
{With a laptop (dual cores, clock speeds 2.6 GHz, memory 16 GB), the computational time of rPCE is shown in Fig.~\ref{timeBoxplot}. Effects of $k$ and $M$ are studied by fixing $M=31$ and $k=10$, respectively. Remark that this figure gives the total time cost for each configuration, i.e., for a specific value of $k$, the total time for constructing $k$ PCE models is given. As observed, since about two thirds of the dataset is used for model construction, the computation time with $k=3$ is only slightly longer than (rather than three times as) with $k=1$. As the value of $k$ increases, since the size of training datasets and the number of model constructions  gets larger, the computational time increases fast. Moreover, in general the computational cost with OMP is higher than that with LARS. Similar phenomena are observed when the value of $M$ increases. When $M=41$, the maximum time cost on running both LARS and OMP is below $18$ minutes which is usually much shorter than the time cost on getting new samples (e.g., $\approx3$ hours are required for getting a new sample in the example of Section \ref{subsec:SAR}).} 

\section{Conclusions}
\label{sec:conclusions}
A new polynomial selection approach, called resampled PCE, has been investigated herein to refine the ranking of importance of candidate polynomials in the context of sparse polynomial chaos expansions. Based on the selected polynomials by LARS and OMP, with the simulation of data variation by resampling, both the selection frequency and the increment on cross-validation error associated with each basis polynomial are arguments in the computation of a total score used in the ranking process. With the PCE model based on rPCE, sensitivity analysis is conveniently performed via the analytical computation of the Sobol' indices based on the expansion coefficients. 

Two factors impact the performance of rPCE. First, the data resampling is conducted by dividing the whole set of data into $k$ similar-sized subsets. The value of $k$ needs to be optimized and set as a combination of good candidates $\{3,5,10,20,N\}$. Second, the candidate polynomials can be generated by LARS, OMP or both. If LARS performs much better than OMP, the resulting selection of polynomials is based on LARS, and vice versa. Otherwise, both the polynomials selected by LARS and OMP would all be treated as candidates in rPCE.   

The performance of rPCE, LARS and OMP is tested on one analytical functions, the  maximum deflection of a truss structure and the estimation of the whole-body SAR (specific absorption rate). In terms of prediction and sensitivity analysis, OMP-based PCE modeling seems the worst among these three methods, especially when the size of ED is small. In contrast, the LARS-based approach generally generates a better model and the refinements by rPCE are obvious in terms of prediction variance and the number of outliers. In any case, rPCE performs as least as well as LARS for global sensitivity analysis.  

Although the size of ED is fixed here, the samples can be automatically enriched to reach a certain accuracy in a specific estimation (e.g., moments) \cite{picheny2010adaptive,blatman2011adaptive,Dubreuil2014,FajraouiMarelli2017}. Moreover, since the building processes with multiple resamples are independent in rPCE, the technique of parallel computations can be applied to ensure the building efficiency of rPCE at the same level with LARS or OMP. 

{In resampled PCE, a high computational cost may be suffered, especially with the suggested setting of $k$, i.e., $k=\{3,5,10,20,N\}$ and a large $N$. As shown by the flow chart in Fig.~\ref{flowChart}, the possible high computational cost is due to the loop about $k$ and $l$. However, one should realize that the loop about $k$ and $l$ are not necessarily performed in sequence and can proceed in parallel. All resampled datasets with different values of $k$ can be first easily generated with LHS method. Since the surrogate modeling with respect to different sets of resamples is separable, techniques of parallel computation (e.g., computation with GPU and distributed computation) can be applied. Moreover, the suggested setting is not compulsory in the running of resampled PCE. Actually, from the results in Section \ref{sec:examples}, we can see the improved modeling performance has been observed with $k=10$, $20$, or $k = \{3,5,10,20\}$, even better performances can be obtained if with the suggested setting. Thus, if the additional computational cost is considered high (especially relative to the cost of obtaining new data), $10$ or $20$ would be suggested for the setting of $k$.}

In forthcoming investigations, more complex scenarios (e.g., electromagnetic dosimetry for human models in the telecommunications network \cite{liorni2015study,kersaudy2015new,huang2017simplified}) are to be analyzed, where a high-order PCE model is often required and the classical approaches easily sink into the overfitting problem. Resampled PCE has the potential to avoid this problem. The refined selection of polynomials reduces the possibility of including redundant or irrelevant basis polynomials in the expansion, thus would have better chances to reach a model with a proper complexity. Here, rPCE combines two forward basis pursuit approaches and the improvements may be slight due to similar selected polynomials. The combination of different kinds of approaches (e.g. forward selection, backward elimination \cite{mantel1970stepdown,blatman2010adaptive} and sparsity-based approach \cite{jakeman2015enhancing}) is open to investigation.

\appendix
\setcounter{table}{0}
\section{Ranking basis polynomials based on LARS or OMP}
\label{appendix:LARSandOMP}
\begin{table*}[!ht]
	\centering
	\caption {Ranking basis polynomials based on orthogonal matching pursuit (OMP).}
	\label{Algorithm_OMP}
	\begin{tabular}{c l}
		\toprule
		1.& Initialization: residual $\bm{R}_0=\bm{y}$, active set $\mathbb{A}_0^a=\varnothing$, candidate set $\mathbb{A}_0^c=\mathbb{A}_{full}$.\\[4pt]
		2.& For $j=1,\ldots,P_{max}=\min\{N-1,\text{card}(\mathbb{A}_{full})\}$, \\
		& \hspace{2pt} 1) Find the basis most correlated with $\bm{R}_{j-1}$,  ${\bm{\alpha}_j}=\arg \max_{\bm{\alpha}\in\mathbb{A}_{j-1}^c} \abs{\bm{R}_{j-1}^T\bm{\psi}_{\bm{\alpha}}}$.\\
		& \hspace{2pt} 2) Update $\mathbb{A}_j^a=\mathbb{A}_{j-1}^a\cup\bm{\alpha}_j$ and $\mathbb{A}_j^c=\mathbb{A}_{j-1}^c\setminus\bm{\alpha}_j$.\\
		& \hspace{2pt} 3) With $\bm{\psi}_{\mathbb{A}_j^a}$, compute $\bm{\beta}_j$ as the OLS solution.\\
		& \hspace{2pt} 4) Update residual $\bm{R}_j=\bm{y}-\bm{\psi}_{\mathbb{A}_j^a}^T\bm{\beta}_j$.\\
		& \small End\\
		\bottomrule
	\end{tabular}
\end{table*}

The PCE model based on orthogonal matching pursuit (OMP) is iteratively built and the iterative procedure is summarized in Table \ref{Algorithm_OMP}. At each iteration, the influence of each polynomial term $\psi_{\bm{\alpha}}$ is measured by its correlation with the data residual $\bm{R}$ (the initial value being $\bm{y}$). The $\bm{\alpha}$ corresponding with the most correlated basis term $\psi_{\bm{\alpha}}$ becomes a member of the active set $\mathbb{A}^a$. Then, computing the basis function $\bm{\psi}_{\mathbb{A}^a}$ supported by the active set, the associated coefficients are obtained by minimizing the least-square error and $\bm{R}$ is updated as the new residual. The most influential polynomials are sequentially selected by repeating the procedure above.

\begin{table}[!ht]
	\centering
	\caption {Ranking basis polynomials based on least angle regression (LARS).}
	\begin{tabular}{c l}
		\toprule
		1. &Initialization: residual $\bm{R}_0=\bm{y}$, active set $\mathbb{A}_0^a=\varnothing$, candidate set $\mathbb{A}_0^c=\mathbb{A}_{full}$;\\[4pt]
		2. &For $j=1,\ldots,P_{max}=\min\{N-1,\text{card}(\mathbb{A}_{full})\}$, \\
		& \hspace{2pt} If $j$ equals 1, define $\bm{u}_1 = \bm{\psi}_{\bm{\alpha}_1}$, $\bm{\alpha}_1=\arg \max_{\bm{\alpha}\in\mathbb{A}_{0}^c} \abs{\bm{R}_{0}^T\bm{\psi}_{\bm{\alpha}}}$, and update $\mathbb{A}_1^a= \{\bm{\alpha}_1\}$, $\mathbb{A}_1^c=\mathbb{A}_{0}^c\setminus\bm{\alpha}_1$.\\
		& \hspace{2pt} Otherwise,\\
		& \hspace{7pt} 1) update $\bm{R}_{j-1}=\bm{R}_{j-2}+\gamma_{j-1}\bm{u}_{j-1}$, $\gamma_{j-1}$ the smallest step length when $\bm{R}_{j-1}$ has the same\\
		& \hspace{7pt} correlation with a {\color{black}basis polynomial} (denoted by $\bm{\psi}_{\bm{\alpha}_j}$, $\bm{\alpha}_j\in \mathbb{A}_{j-1}^c$) as those with all {\color{black}polynomials}\\
		& \hspace{7pt} in $\bm{\psi}_{\mathbb{A}_{j-1}^a}$.\\
		& \hspace{7pt} 2) update $\mathbb{A}_j^a=\mathbb{A}_{j-1}^a\cup\bm{\alpha}_j$ and $\mathbb{A}_j^c=\mathbb{A}_{j-1}^c\setminus\bm{\alpha}_j$. \\
		& \hspace{7pt} 3) compute the equiangular vector of all {\color{black}polynomials} in  $\bm{\psi}_{\mathbb{A}_j^a}$ as  $\bm{u}_{j}$.\\
		&End \\
		\bottomrule
	\end{tabular}
	\label{Algorithm_LARS}
\end{table}

Least angle regression (LARS) is a less greedy version of traditional forward selection methods. It is known that different flavors of LARS yield efficient solutions of LASSO \cite{tibshirani1996regression} (which constrains both the data discrepancy by ordinary least square and the sparsity of regression coefficients by $\ell_1$-norm) and forward stagewise linear regression \cite{weisberg2005applied} (another promising model-selection method), respectively.

The iterative algorithm of sparse PCE modeling based on LARS (originally proposed in \cite{blatman2011adaptive}) is given in Table \ref{Algorithm_LARS}, where details on how to compute step length $\gamma_{j-1}$ and equiangular vector $\bm{u}_j$ can be found in \cite{efron2004least}. As seen from this short summary, the building process is similar with the one based on OMP, except that from the second iteration since the residual $\bm{R}$ evolves along the equiangular directions of basis functions other than along basis functions themselves.

\section*{Acknowledgments}
The first author has been supported in part by the Emergence programme of the Science and Technologies of Information and Communication (STIC) department, University Paris-Saclay.


\end{document}